\documentclass[a4paper,11pt]{article}
\pdfoutput=1 

\usepackage{jheppub}

\usepackage{graphicx, epstopdf}

\usepackage[utf8]{inputenc}

\usepackage{amsmath, amsthm, amsfonts, amssymb}
\usepackage{mathtools, mathabx, bm, bbm, scalerel, xfrac}
\usepackage{physics, slashed}
\usepackage{hyperref}
\usepackage{color}
\usepackage{xspace} 

\allowdisplaybreaks

\title{Complete collection of one-loop triple-collinear splitting operators for dimensionally-regulated QCD}

\author[a]{Micha\l{} Czakon}
\affiliation[a]{Institut f\"ur Theoretische Teilchenphysik und Kosmologie, RWTH Aachen University,\\ D-52056 Aachen, Germany}
\emailAdd{mczakon@physik.rwth-aachen.de}
\author[b]{and Sebastian Sapeta}
\affiliation[b]{Institute of Nuclear Physics, Polish Academy of Sciences, Radzikowskiego 152,\\ 31-342 Krak\'ow, Poland}
\emailAdd{sebastian.sapeta@ifj.edu.pl}

\abstract{We provide results for the one-loop triple-collinear color/spin-space splitting operators for the five possible processes, $q \to qq'\bar{q}'$, $q \to qq\bar{q}$, $q \to qgg$, $g \to gq\bar{q}$ and $g \to ggg$. The expressions are exact in dimensionally-regulated massless QCD up to a single integral, which we expand to second order in the dimensional-regularisation parameter. We also evaluate the related splitting functions. Our results are both sufficient and indispensable for the construction of subtraction and integrated-subtraction terms for triple-collinear singularities of one-loop double-real-emission cross-section contributions as part of a next-to-next-to-next-to leading order subtraction scheme.}

\preprint{P3H-22-042, TTK-22-15, IFJPAN-IV-2022-7}

\begin{document}
\maketitle
\flushbottom

\section{Introduction}

The study of the singular behaviour of massless-gauge-theory scattering amplitudes in multiparticle soft/collinear limits has a long history. One of the several important applications of the results of the analyses is the construction of efficient methods for the evaluation of cross sections in perturbative QCD defined within dimensional regularisation. Of particular interest for the present work are subtraction schemes, see e.g.\ Refs.~\cite{Catani:1996vz,Frixione:1995ms} for classic work at the next-to leading order (NLO). The quest for predictions at ever higher orders of perturbation theory has led to the construction of a plethora of schemes. The current frontier lies between the next-to-next-to (NNLO) and next-to-next-to-next-to (N3LO) order, see the review Ref.~\cite{Caola:2022ayt} and references therein.

The known singular limits of scattering amplitudes can be classified by the order of perturbation theory of the involved matrix elements, and by the number of unresolved partons whose presence cannot be detected either because they have vanishingly small (soft) energies or their momenta are parallel (collinear) to the momentum of a single parton. A complete set of formulae at tree-level for up to three unresolved partons has been amassed in Refs.~\cite{campbell,Catani:1998nv,Catani:1999ss,DelDuca:1999iql,Kosower:2002su,Catani:2019nqv,DelDuca:2019ggv,DelDuca:2020vst,Braun-White:2022rtg}. At the one-loop level, complete information is available for a single unresolved parton \cite{Bern:1994zx,Bern:1998sc,Kosower:1999rx,Bern:1999ry,Catani:2000pi}, while the case of two unresolved partons has attracted much attention \cite{Catani:2003vu,Sborlini:2014mpa,Sborlini:2014kla,Badger:2015cxa,Zhu:2020ftr,Catani:2021kcy}. Finally, at the two-loop level only single unresolved limits have been analysed extensively \cite{Bern:2004cz,Badger:2004uk,Duhr:2014nda,Li:2013lsa,Duhr:2013msa}.

In the present publication, we are concerned with the triple-collinear limit of one-loop amplitudes. It is surprising that the known results for this case do not even cover all possible splitting processes. While Ref.~\cite{Badger:2015cxa} provides results for a gluon splitting into three partons, the much earlier Ref.~\cite{Catani:2003vu} contains the case of a quark splitting into a quark and a quark-anti-quark pair of different flavor, but the result is incomplete\footnote{The authors provide the antisymmetric part of the results with respect to the exchange of the quark and the anti-quark momenta. The tree-level splitting is symmetric under this transformation.}. The previously cited Refs.~\cite{Sborlini:2014mpa,Sborlini:2014kla} are rather concerned with processes involving a photon. Thus, results for quark splittings, $q \to qq'\bar{q}'$, $q \to qq\bar{q}$, $q \to qgg$, are either incomplete or not available. In view of this situation, the following quote from the almost twenty years old Ref.~\cite{Catani:2003vu} is truly ironic:
\vspace{.5cm}
\\
\noindent
\textit{ "For the sake of brevity, we have limited ourselves, in this letter, to presenting a few explicit results for the one-loop triple collinear splitting. These results have mainly an illustrative purpose. The method and the tools (in particular, the one-loop integrals) used to obtain them are sufficient and can be applied straightforwardly to evaluate the one-loop splitting matrix of any splitting process $a \to a_1 + a_2 + a_3$."}
\vspace{.5cm}
\\
There is yet another problem with the available results for the one-loop triple-collinear splittings. They are restricted to an expansion in the dimensional-regularisation parameter up to finite terms. Below, we explain that this is insufficient for an application to the construction of a subtraction scheme. This fact was, unsurprisingly, known to at least the authors of Ref.~\cite{Badger:2015cxa} as can be read in the conclusions to that work.

In the present publication, we resolve the aforementioned issues and provide the complete set of results for triple-collinear splittings, $q \to qq'\bar{q}'$, $q \to qq\bar{q}$, $q \to qgg$, $g \to gq\bar{q}$ and $g \to ggg$, in the form necessary for the construction of an N3LO subtraction scheme.

We aim at a self-contained but concise publication. In the first Section, we provide the necessary definitions. Subsequently, we present the methods used in the calculation, and explain the requirements for the construction of a subtraction scheme in relation to the results for the triple-collinear splittings. In the third Section, we discuss the evaluation of the occurring Feynman integrals and provide several new results. After a short outlook on future work, we provide more infomation on the special functions involved in this study in two Appendices. The results obtained are lengthy. We believe that it is crucial to provide them in an easily accessible form, and there is no better form than a set of ancillary files that can be easily manipulated. We use the format of the computer algebra system \textsc{Mathematica} \cite{Mathematica} and describe the files in the last Appendix.

\section{Splitting operators and splitting functions}

\subsection{Definitions and properties}

We consider QCD with $n_l$ massless quark fields, defined through perturbation theory supplemented with conventional dimensional regularisation\footnote{Our results assume that gluons have $d-2$ polarisation states.} with spacetime-dimension parameter $d = 4-2\epsilon$ and dimension-setting scale $\mu$. Any parton-scattering amplitude viewed as a vector in color- and spin-space (see Ref.~\cite{Catani:1996vz} for a pedagogical introduction to the formalism) may be expanded in the bare coupling constant, $g_s^B$:
\begin{equation}
    \ket{M} \equiv ( \mu^{-\epsilon} g_s^B )^n \Big( \ket{M^{(0)}} + \frac{\mu^{-2\epsilon}\alpha^B_s}{(4\pi)^{1-\epsilon}} \, \ket{M^{(1)}} + \order{\alpha_s^2} \Big) \; , \qquad \alpha^B_s \equiv \frac{(g^B_s)^2}{4\pi} \; .
\end{equation}
Renormalisation is not essential in the present context, since splitting operators that are the topic of this work renormalise as ordinary amplitudes. Nevertheless, we use $\mu^{-2\epsilon}\alpha^B_s/(4\pi)^{1-\epsilon}$ as expansion parameter as it is dimensionless and thus yields fixed-order amplitudes $\ket{M^{(n)}}$ of integer mass dimension, because $\ket{M}$ has this property. Furthermore, there are no $\ln(4\pi)$ terms in the results as in the $\overline{\rm{MS}}$ renormalisation scheme.

Consider now an amplitude's triple-collinear limit for partons $f_1,f_2,f_3$, where $f_i$ may denote a gluon, a quark or an anti-quark, with outgoing momenta $p_i$, $i=1,2,3$. The case of some momenta incoming can be inferred by crossing and analytic continuation of the Feynman integrals. Our results are provided for a final state splitting for definiteness. In the present context, it is sufficient to consider two quark flavors denoted generically by $q$ and $q'$. The amplitude is known to factorize at leading power in $s_{123} \equiv p_{123}^2 \equiv (p_1+p_2+p_3)^2$ as discussed for the particular case considered here in Ref.~\cite{Catani:2003vu} (see also Ref.~\cite{Feige:2014wja} for a pedagogical general discussion)\footnote{We use the standard notation for asymptotic expansions throughout this publication.}:
\begin{equation} \label{eq:factorisation}
    \ket{M_{f_1f_2f_3\dots}(p_1,p_2,p_3,\dots)} \quad \sim \quad  \textbf{Split}_{f_1f_2f_3}(p_1,p_2,p_3) \, \ket{M_{f\dots}(p,\dots)} \qquad \big( s_{123} \to 0 ) \; .
\end{equation}
$\textbf{Split}_{f_1f_2f_3}(p_1,p_2,p_3)$ is the splitting operator that we wish to evaluate in a perturbative expansion up to $\order{\alpha_s^B}$:
\begin{equation}
    \textbf{Split}_{f_1f_2f_3}(p_1,p_2,p_3) = \big( \mu^{-\epsilon}g^B_s \big)^2 \, \Big( \textbf{Split}^{(0)}_{f_1f_2f_3}(p_1,p_2,p_3) + \frac{\mu^{-2\epsilon}\alpha^B_s}{(4\pi)^{1-\epsilon}} \,  \textbf{Split}^{(1)}_{f_1f_2f_3}(p_1,p_2,p_3) + \order{(\alpha_s^B)^2} \Big) \; .
\end{equation}
The splitting operator acts on the color and polarisation indices of the parton $f$ in $\ket{M_{f\dots}(p,\dots)}$ and yields a vector in color- and spin-space of the nearly collinear partons $f_{1,2,3}$. The splitting parton $f$ is determined from $f_{1,2,3}$ by flavor conservation, while the factorization is only valid if $f$ is well defined. The momentum, $p$, of the splitting parton $f$ is defined with the help of an auxiliary light-like vector $q$ as follows:
\begin{equation}
    p_{123} \equiv p + \frac{s_{123}}{2\,( p_{123} \cdot q )} \, q \; , \qquad p^2 = q^2 = 0 \; , \quad p \cdot q \neq 0 \; , \quad p^0,q^0 > 0 \; .
\end{equation}
Suitable $p$ and $q$ may always be uniquely constructed up to normalisation of $q$, assuming that the three-vectors, $\bm{q}$ and $\bm{p}$, are parallel, $\bm{q} \parallel \bm{p} \parallel \bm{p}_{123}$. Since the singularity structure of one-loop QCD amplitudes is completely known \cite{Catani:2000ef}, it is possible to derive the singularity structure of the one-loop splitting operators \cite{Catani:2003vu}:
\begin{equation} \label{eq:Ioperator}
\begin{split}
    \textbf{Split}&^{(1)}_{f_1f_2f_3}(p_1,p_2,p_3) = \\[.2cm]& \frac{\Gamma(1+\epsilon)\Gamma^2(1-\epsilon)}{\Gamma(1-2\epsilon)} \Bigg[ \frac{2}{\epsilon^2} \sum_{1 \leq i < j \leq 3} \textbf{T}_i \cdot \textbf{T}_j \, \Bigg( \frac{-2 (p_i \cdot p_j) - i0^+}{\mu^2} \Bigg)^{-\epsilon} \\[.2cm]&\qquad\qquad\qquad\qquad
    +\Bigg( \frac{-s_{123}-i0^+}{\mu^2} \Bigg)^{-\epsilon} \Bigg( \frac{2}{\epsilon^2} \sum_{1 \leq i \leq j \leq 3} \textbf{T}_i \cdot \textbf{T}_j \, \Big( 2 - \Big( \frac{p_i \cdot q}{p_{123} \cdot q} \Big)^{-\epsilon} - \Big( \frac{p_j \cdot q}{p_{123} \cdot q} \Big)^{-\epsilon}\Big) \\[.2cm]&\qquad\qquad\qquad\qquad\qquad\qquad\qquad\qquad\qquad
    +\frac{1}{\epsilon} \Big( b_0 + \gamma_f - \sum_{1 \leq i \leq 3} \gamma_i \Big) \Bigg) \Bigg] \, \textbf{Split}^{(0)}_{f_1f_2f_3}(p_1,p_2,p_3)
    \\[.2cm]& + \order{\epsilon^0} \; ,
\end{split}
\end{equation}
where $\textbf{T}_i$ are color-space operators corresponding to the generators, $T^a_{c'c}$, of the fundamental representation for quarks, the generators, $-T^{a\, \star}_{c'c}$, of the anti-fundamental representation for anti-quarks, and the generators, $if^{c'ac}$, of the adjoint representation for gluons. Furthermore, $\gamma_q = \gamma_{\bar{q}} = 3/2C_F$, $\gamma_g = b_0/2$, $b_0 = (11/3C_A - 4/3T_Fn_l)$, $C_F = (N_c^2-1)/2N_c = 4/3$, $C_A = N_c = 3$. The presence of the leading coefficient of the QCD $\beta$-function, $b_0$, is due to the lack of renormalisation. We stress that, in the above formula, all terms of $\order{\epsilon^0}$ are arbitrary. For convenience of the reader, we provide the expressions obtained from Eq.~\eqref{eq:Ioperator} for each of the splitting operators in an ancillary file attached to this publication. There, we only keep the pure-pole contributions, proportional to $1/\epsilon^2$ and $1/\epsilon$, in the pre-factor in Eq.~\eqref{eq:Ioperator} acting on the tree-level splitting operators $\textbf{Split}^{(0)}_{f_1f_2f_3}(p_1,p_2,p_3)$.

\begin{figure}[h]
    \centering
    \includegraphics[width=.85\textwidth]{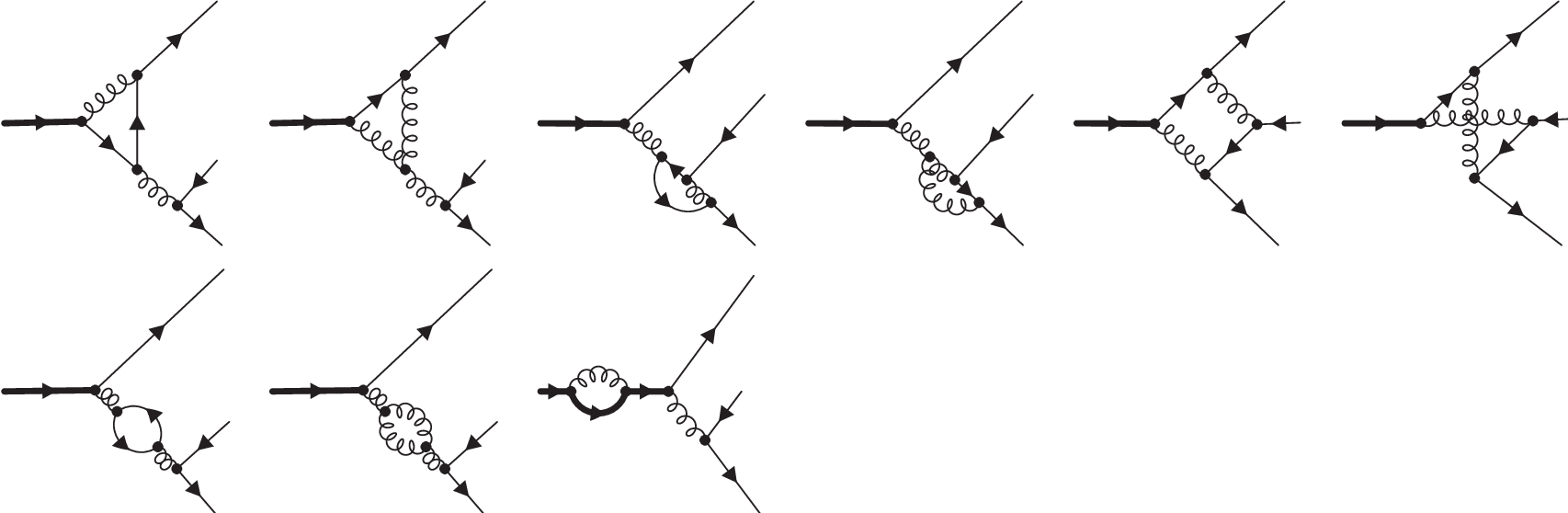}
    \caption{9 diagrams contributing to the splitting operator $q \to qq'\bar{q}'$. The thick incoming lines on the left side of each diagram correspond to the off-shell splitting parton. The outgoing lines on the right side of each diagram are on shell. The anti-quark line carries an arrow of direction opposite to that of the energy flow.}
    \label{fig:qqq}
\end{figure}
\begin{figure}[h]
    \centering
    \includegraphics[width=.85\textwidth]{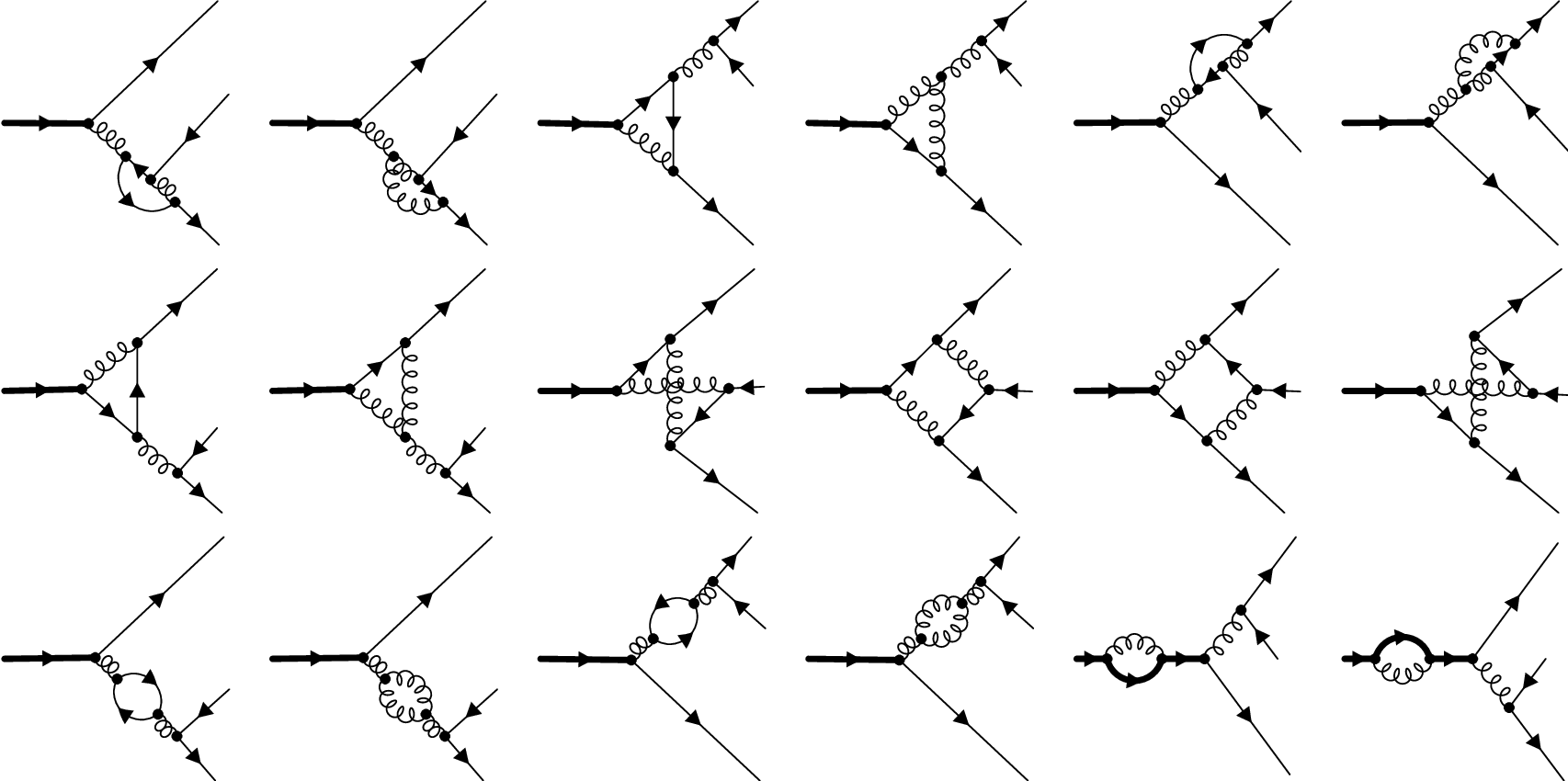}
    \caption{18 diagrams contributing to the splitting operator $q \to qq\bar{q}$. Description as in Fig.~\ref{fig:qqq}.}
    \label{fig:qqqid}
\end{figure}
\begin{figure}[h]
    \centering
    \includegraphics[width=.85\textwidth]{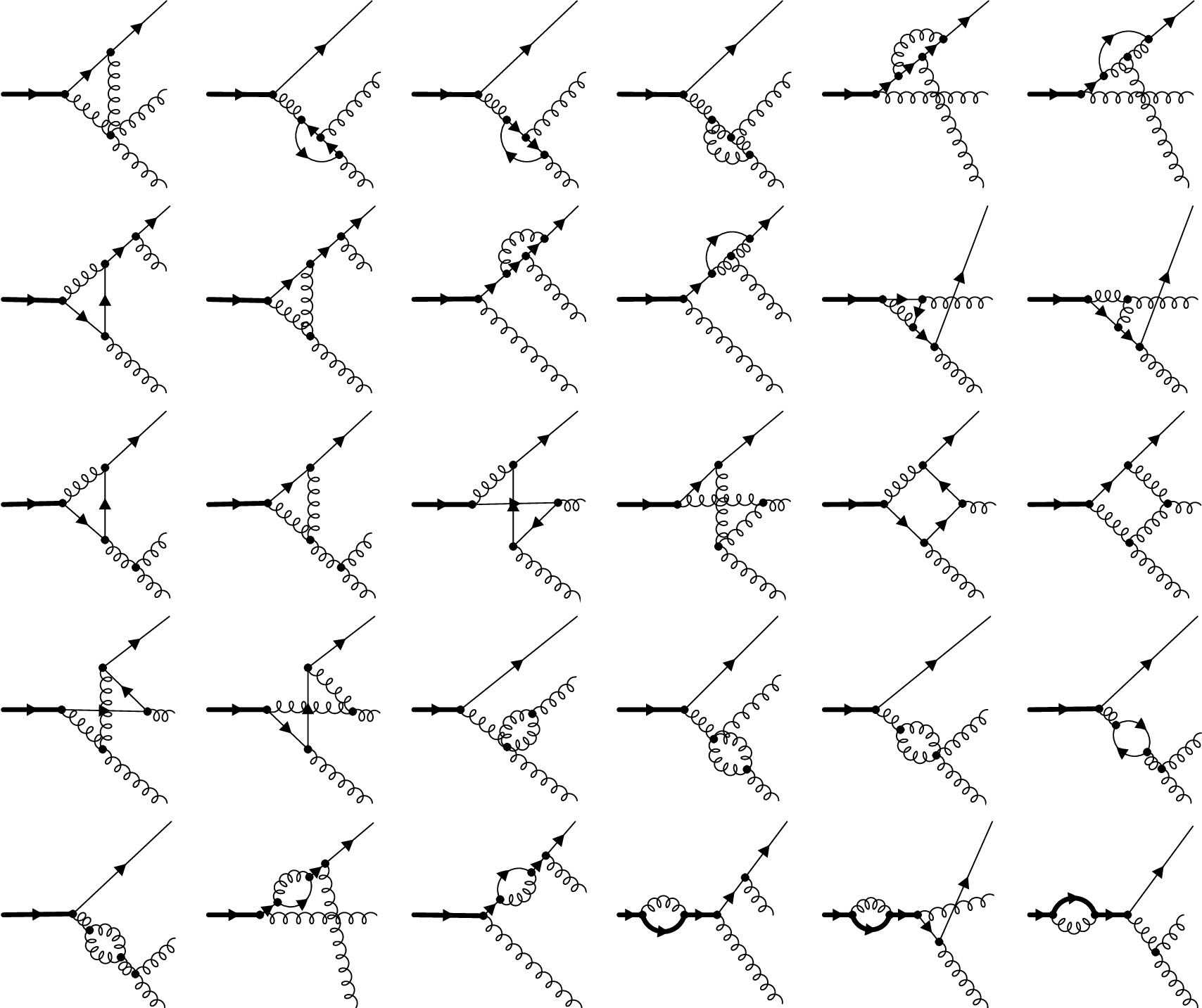}
    \caption{30 diagrams contributing to the splitting operator $q \to qgg$ Description as in Fig.~\ref{fig:qqq}.}
    \label{fig:qgg}
\end{figure}
\begin{figure}[h]
    \centering
    \includegraphics[width=.85\textwidth]{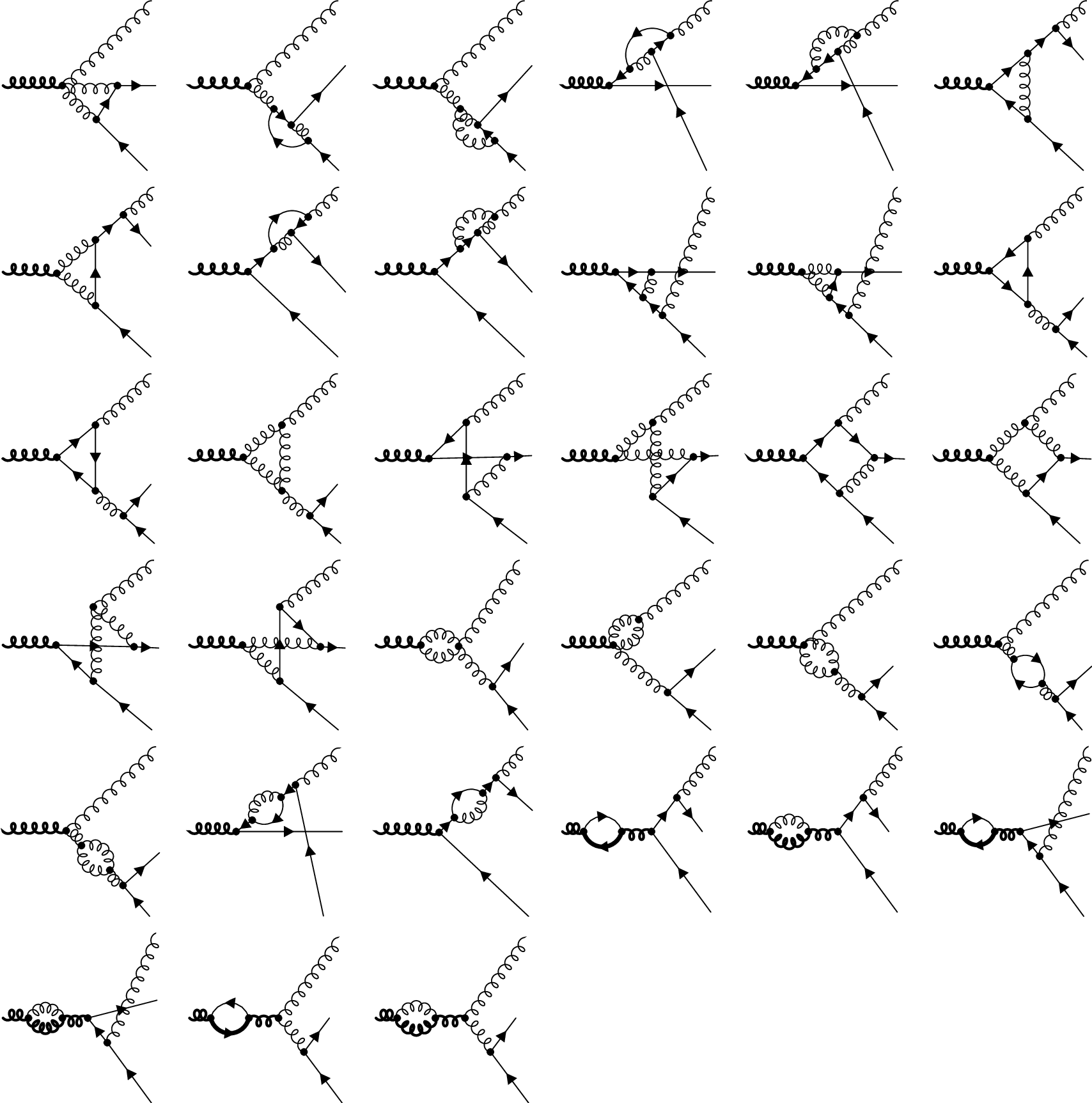}
    \caption{33 diagrams contributing to the splitting operator $g \to gq\bar{q}$. Description as in Fig.~\ref{fig:qqq}.}
    \label{fig:gqq}
\end{figure}
\begin{figure}[h]
    \centering
    \includegraphics[width=.85\textwidth]{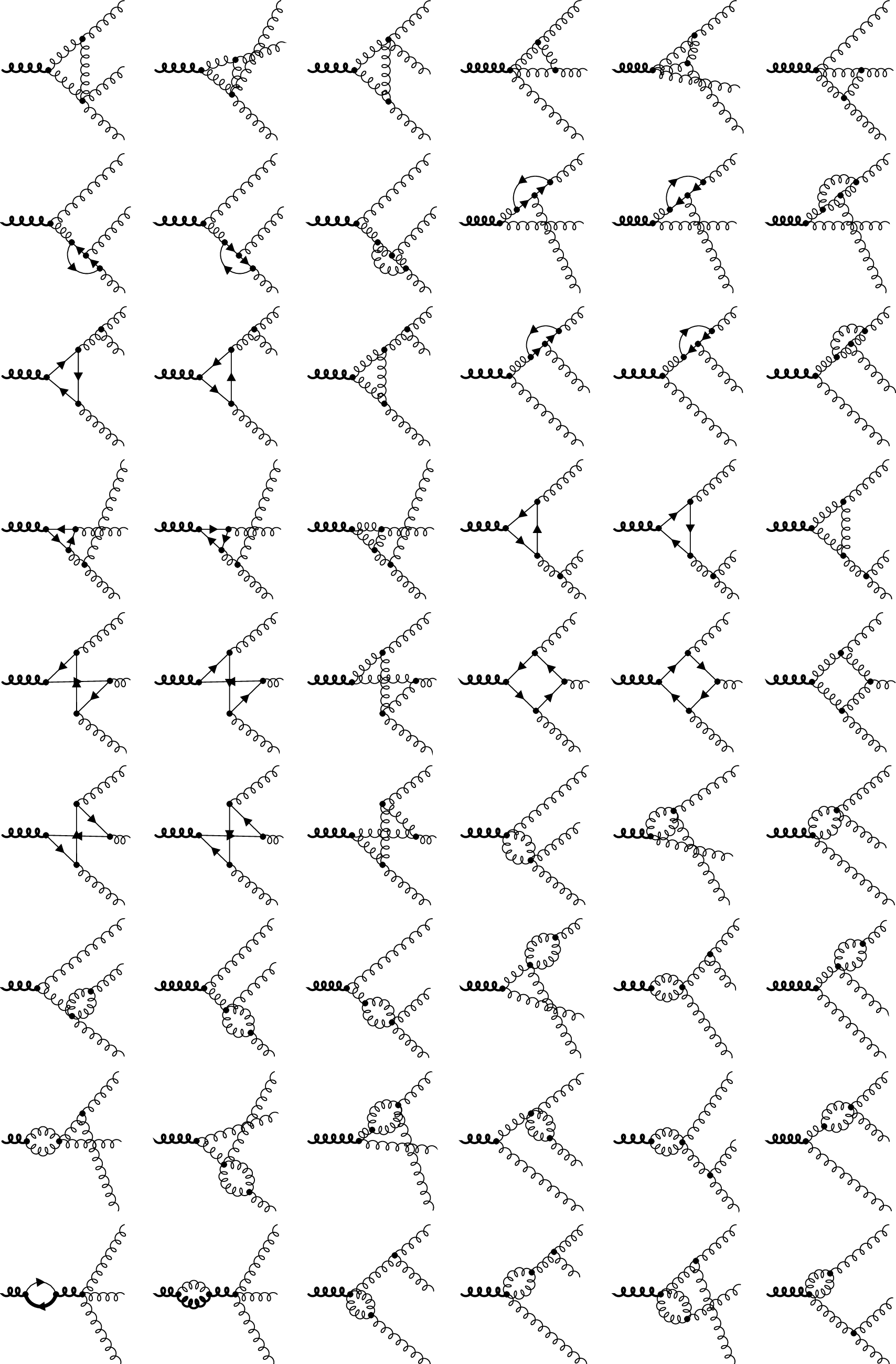}
    \caption{Part of the 68 diagrams contributing to the splitting operator $g \to ggg$. Description as in Fig.~\ref{fig:qqq}.}
    \label{fig:ggg1}
\end{figure}
\begin{figure}[h]
    \centering
    \includegraphics[width=.85\textwidth]{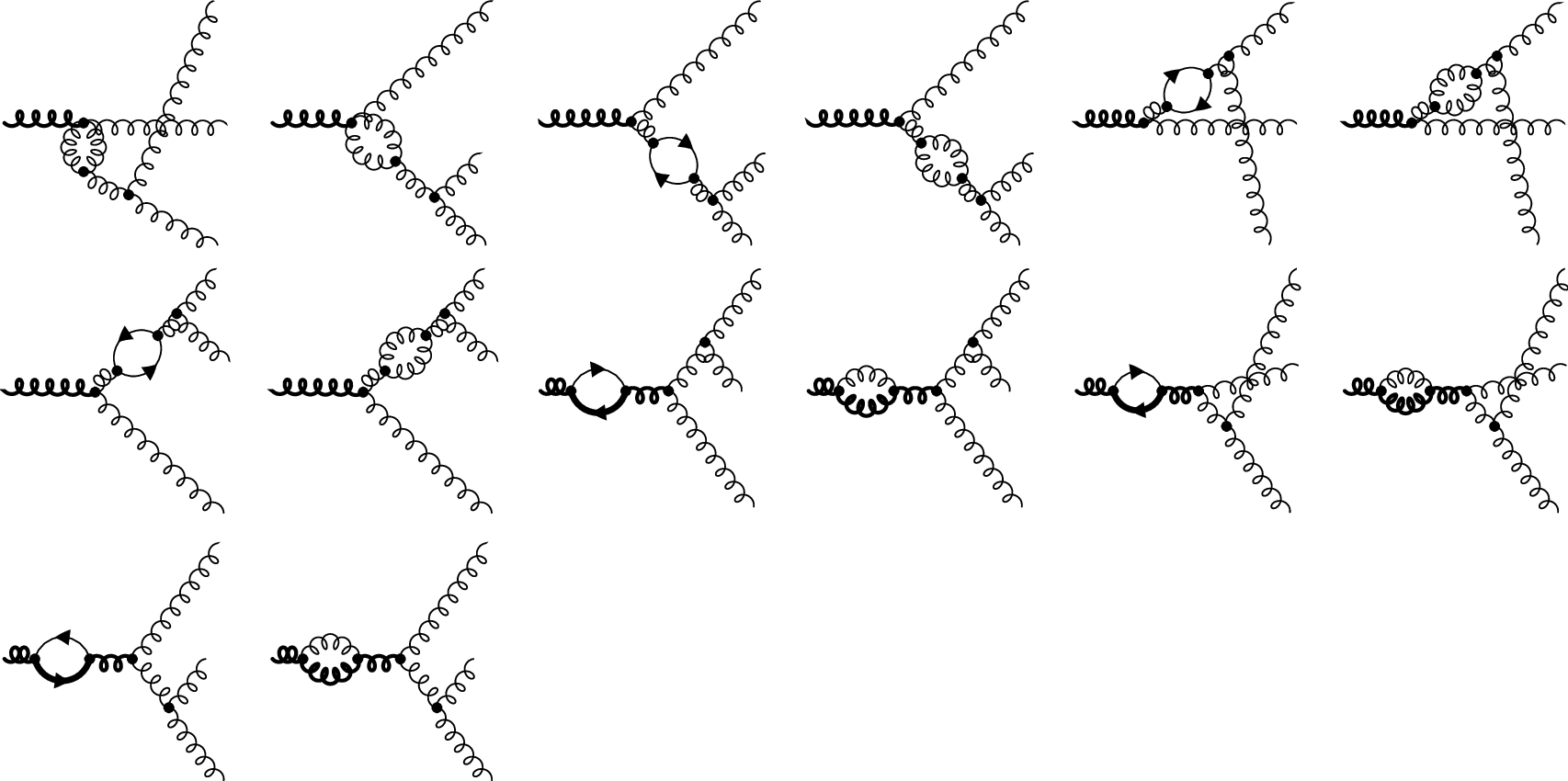}
    \caption{Part of the 68 diagrams contributing to the splitting operator $g \to ggg$. Description as in Fig.~\ref{fig:qqq}.}
    \label{fig:ggg2}
\end{figure}

The splitting operators for the five possible splittings\footnote{Anti-quark splitting operators are related to those of quarks by charge-conjugation symmetry of QCD.}, $q \to qq'\bar{q}'$, $q \to qq\bar{q}$, $q \to qgg$, $g \to gq\bar{q}$ and $g \to ggg$ are obtained from diagrams depicted in Figs.~\ref{fig:qqq}, \ref{fig:qqqid}, \ref{fig:qgg}, \ref{fig:gqq} and \ref{fig:ggg1}, \ref{fig:ggg2} respectively. The incoming off-shell line of the splitting parton is contracted with a massless Dirac spinor $u_0 \equiv u(p,\lambda_0)$ for a quark splitting, and with a massless transverse polarisation vector $\epsilon_0 \equiv \epsilon(p,\lambda_0)$ for a gluon splitting, with $\lambda_0$ the helicity of the splitting parton. This prescription obviously realises Eq.~\eqref{eq:factorisation} by approximating the numerator of the splitting parton's propagator by a polarisation sum of a product of on-shell spinor/polarisation vectors. The only non-trivial issue is the necessity of using a light-cone-gauge gluon propagator:
\begin{equation}
    \ev{\tilde{A}_\mu^a(k) A_\nu^b(0)}{0} = \frac{i\delta^{ab}}{k^2+i0^+} \Big( - g_{\mu\nu} + \frac{q^\mu k^\nu + q^\nu k^\mu}{q \cdot k}\Big) \; ,
\end{equation}
where $A_\mu^a$ is the free gluon field, tilde denotes Fourier transformation, and $q$, $q^2 = 0$ is an auxilliary light-like vector. This is, of course, part of the proof of the factorisation theorem. For more details we refer the reader again to Ref.~\cite{Feige:2014wja}.

Besides splitting operators, one also defines dimensionless splitting functions and averaged splitting functions:
\begin{equation} \label{eq:SplittingFunctions}
    \bm{\hat{P}}_{f_1f_2f_3} \equiv \Big( \tfrac{1}{2} s_{123} \Big)^2 \, \textbf{Split}^\dagger_{f_1f_2f_3} \textbf{Split}_{f_1f_2f_3}  \; , \qquad \ev{\hat{P}_{f_1f_2f_3}} \equiv \frac{1}{n_f^c n^s_f} \Tr(\bm{\hat{P}}_{f_1f_2f_3}) \; ,
\end{equation}
where $n_f^c$ and $n_f^s$ are the numbers of color and spin degrees-of-freedom of the splitting parton, $n^c_q = 3$, $n^c_g = 8$, $n^s_q = 2$ and $n^s_g = d-2$. The color- and spin-space operators $\bm{\hat{P}}_{f_1f_2f_3}$ are relevant for the study of the triple-collinear limit of amplitudes squared and summed over colors and polarisations of the external states. By color conservation, they are proportional to the identity operator in color space. Due to helicity conservation in massless QCD, the same holds for spin space in the case of quarks. The corresponding averages $\ev{\hat{P}_{f_1f_2f_3}}$ are necessary for the determination of the integrated subtraction terms in the construction of subtraction schemes, see Section~\ref{Sec:Subtraction}. As a result of our work, we provide:
\begin{equation} \label{eq:SplittingFunctions1l}
    \bm{\hat{P}}^{(1)}_{f_1f_2f_3} \equiv \Big( \tfrac{1}{2} s_{123} \Big)^2 \, \Big(    \textbf{Split}^{(0)\,\dagger}_{f_1f_2f_3} \textbf{Split}^{(1)}_{f_1f_2f_3} + \text{hermitian conjugate} \Big) \; ,
\end{equation}
as well as its averaged version.

\subsection{Evaluation of the diagrams} \label{Sec:Diagrams}

The calculation proceeds in standard fashion. The diagrams are generated with private software (although the figures have been produced with \textsc{FeynArts} \cite{Hahn:2000kx}) and simplified with the program \textsc{FORM} \cite{Vermaseren:2000nd}. Subsequently, we perform a Passarino-Veltman \cite{Passarino:1978jh} reduction of the tensor integrals, consisting of expressing integrals with numerators proportional to $l^{\mu_1} \cdots l^{\mu_n}$, with $l$ the loop momentum and $n$ the tensor rank, as sums of symmetric tensors built of products of $p_{1,2,3}$, $q$ and the metric tensor $g$, multiplied with scalar integrals with each $l^{\mu_i}$ contracted with another momentum. The reduction formulae are efficiently derived with the help of the program \textsc{Fermat} \cite{fermat}. In the case of Feynman integrals whose denominators contain up to three different momenta from the set $\{p_1,p_2,p_3,q\}$, we encounter tensors of rank up to four. In the case of integrals whose denominators contain all four momenta, the maximal tensor rank is two. Since four linearly-independent vectors form a basis in four-dimensional space, the metric tensor itself can be expressed as a linear combination of products of two out of the four momenta. Hence, the procedure is singular in four dimensions. As we are working in dimensional regularisation, this leads to the appearance of a pole, $1/\epsilon$, in the coefficients of the Passarino-Veltman reduction. It has been noticed in the study of ordinary (i.e.\ without linear propagators) tensor pentagon integrals in Ref.~\cite{Bern:1993kr} that these poles are removed by switching from four-dimensional to six-dimensional pentagon integrals. The same phenomenon occurs here as well, see Section~\ref{Sec:LCGBox}. After Passarino-Veltman reduction, we perform a further integration-by-parts reduction using the software package \textsc{Kira} \cite{Maierhofer:2017gsa, Klappert:2020nbg} for convenience, since it allows for linear propagators. The number of occurring integrals is reduced to thirty four, see Section~\ref{Sec:Masters}. The reduction introduces spurious $1/\epsilon$ poles in coefficients of the bubble integrals Eq.~\eqref{eq:Bubble1} and \eqref{eq:Bubble2}. This is the price for the removal of triangle integrals, i.e.\ integrals with three ordinary Feynman propagators. Since the bubble integrals are known exactly, these spurious poles do not constitute a difficulty.

At this point, the results for the splitting operators still have a complicated color and spin structure. Both of these structures can be simplified by the following two algorithms. As far as color factors are concerned, we use the formulae:
\begin{equation} \label{eq:Cvitanovic}
    if^{abc} = \frac{1}{T_F} \Tr(\comm{T^a}{T^b}T^c) \; , \qquad T^a_{ij}T^a_{kl} = T_F \Big( \delta_{il}\delta_{kj} - \frac{1}{N_c} \delta_{ij} \delta_{kl} \Big) \; , \qquad T_F = \frac{1}{2} \; , \quad N_c = 3 \; .
\end{equation}
This procedure is commonly referred to as the Cvitanovic algorithm \cite{Cvitanovic:1976am}. For definiteness, we assign the following color indices: $c_0$ to the splitting parton, and $c_i$, $i = 1,2,3$ to the outgoing partons. This algorithm provides a unique basis of color structures. Below, we list the occuring ones for each of the splitting functions. 

The spin structure is simplified in several steps. First, $\slashed{q}$ is shifted to the right in any product of Dirac $\gamma$-matrices using the Dirac algebra. If possible, this step is followed by the application of:
\begin{equation}
    \slashed{q} \, u_0 = \frac{2 (p_{123} \cdot q)}{p_{123}^2} \, \slashed{p}_{123} \, u_0 \; .
\end{equation}
Subsequently, equations of motion for spinors $\bar{u}_i \equiv \bar{u}(p_i,\lambda_i)$, $i=1,2$, $v_3 \equiv v(p_3,\lambda_3)$:
\begin{equation}
  \bar{u}_1 \, \slashed{p}_1 = \bar{u}_2 \, \slashed{p}_2 = \slashed{p}_3 v_3 = 0 \; ,
\end{equation}
are used, once $\slashed{p}_{1,2}$ has been shifted to the left, and $\slashed{p}_3$ to the right in any product of $\gamma$-matrices. Polarisation vectors $\slashed{\epsilon}_{0}$ and $\slashed{\epsilon}^*_{1,2}$ are shifted to the left. The assumed transversality of the polarisation vectors to not only the respective momentum, but also to $q$:
\begin{equation}
    p \cdot \epsilon_0 = q \cdot \epsilon_0 = p_i \cdot \epsilon^*_i = q \cdot \epsilon^*_i = 0 \; , \qquad i = 1,2,3 \; ,
\end{equation}
is used at each step together with the implied relation:
\begin{equation}
    p_3 \cdot \epsilon_0 = -(p_1+p_2) \cdot \epsilon_0 \; .
\end{equation}
This last relation breaks the explicit permutation symmetry in the case of the $g \to ggg$ splitting, but is necessary to obtain a basis of spin-space structures without redundancy.

With the above, the expressions for the splitting operators attain their final unique representation in dimensional regularisation. The spin structures are, however, not linearly independent in four dimensions. In Ref.~\cite{Bern:1993kr}, it was pointed out that six-dimensional scalar pentagon integrals disappear from the finite $\order{\epsilon^0}$ terms of the $\epsilon$-expansion of scattering amplitudes, if four-dimensional relations between spin structures are exploited. The same happens to hold in our case for six-dimensional integrals containing four ordinary and one linear propagator discussed in Section~\ref{Sec:LCGBox}. In order to make this fact explicit, we represent any vector, including $\gamma$-matrices in a basis of the four vectors $k_i \equiv p_i$, $i = 1,2,3$, $k_4 \equiv q$:
\begin{equation} \label{eq:4Dprojection}
    v^\mu = \sum_{i,j=1}^4 k_i^\mu \, \big( K^{-1} \big)_{ij} \, k_j \cdot v \; , \qquad K_{ij} \equiv k_i \cdot k_j \; .
\end{equation}
The resulting splitting operators are only valid in four dimensions. On the other hand, the remaining spinor chains in the case of splittings involving quarks are vastly simplified and can be trivially expressed in the popular spinor-helicity formalism using $v_{3\pm} \; \propto \; u_{3\mp}$ (up to a phase factor) and with $i,k \in \{1,2,3\}$, $j \in \{0,\dots,3\}$:
\begin{equation}
\begin{gathered}
    \bar{u}_{i\pm} \, \slashed{p}_k \, u_{j\pm} = \big( \bar{u}_{i\pm}u_{k\mp} \big) \, \big( \bar{u}_{k\mp}u_{j\pm} \big) \; , \qquad \bar{u}_{i\pm} \, \slashed{q} \, u_{j\pm} = 2\sqrt{(p_i \cdot q)(p_j \cdot q)} \; , \\[.2cm]
    p_i \cdot \epsilon^*_{j\pm} = \pm \frac{\big( \bar{u}_{j\pm}u_{i\mp} \big) \, \big( \bar{u}_{i\mp}u_\pm(q) \big)}{\sqrt{2}\bar{u}_\mp(q)u_{j\pm}} \; ,
\end{gathered}
\end{equation}
where the subscripts $\pm$ denote helicity, while the unlisted cases vanish. As long as there are spinor chains, this procedure does not yield larger expressions than those valid in general $d$ dimensions. As splitting operators expanded up to $\order{\epsilon^0}$ are useful to subtract the singularities of exact amplitudes in practical calculations, we also provide them as ancillary files attached to this publication, just as we do with for the exact results which are far too lengthy to be useful in printed text.

In order to illustrate the structure of the splitting operators, we now list the occurring color and spin structures for each of them.

\subsubsection*{\underline{$q \to q_1q'_2\bar{q}'_3$}}

In this case, there are two color structures both at tree- and one-loop level:
\begin{equation}
    \delta_{c_1c_0}\delta_{c_2c_3} \; , \qquad \delta_{c_2c_0}\delta_{c_1c_3} \; . 
\end{equation}
At tree level, nevertheless, they are actually generated from a single diagram proportional to $T^a_{c_1c_0}\,T^a_{c_2c_3}$, which is split into two terms by the Cvitanovic algorithm according to Eqs.~\eqref{eq:Cvitanovic}.
Furthermore, there are only three distinct spin structures at tree level:
\begin{equation}
\,\bar{u}_1\,\gamma^{\mu}\,u_0\,\bar{u}_2\,\gamma_{\mu}\,v_3\,\; ,
\qquad \,\bar{u}_1\,\slashed{p}_2\,u_0\,\bar{u}_2\,\slashed{q}\,v_3\,\; ,
\qquad \,\bar{u}_1\,\slashed{p}_3\,u_0\,\bar{u}_2\,\slashed{q}\,v_3\,\; ,  \label{eq:spin0lqQQ}
\end{equation}
while there appear the following additional structures at one-loop level:
\begin{gather}
\,\bar{u}_1\,\gamma^{\mu_1}\,\gamma^{\mu_2}\,\gamma^{\mu_3}\,u_0\,\bar{u}_2\,\gamma_{\mu_1}\,\gamma_{\mu_2}\,\gamma_{\mu_3}\,v_3\,\; ,
\qquad \,\bar{u}_1\,\gamma^{\mu}\,u_0\,\bar{u}_2\,\slashed{p}_1\,\gamma_{\mu}\,\slashed{q}\,v_3\,\; ,
\qquad \,\bar{u}_1\,\gamma^{\nu_1}\,\gamma^{\nu_2}\,\slashed{p}_3\,u_0\,\bar{u}_2\,\gamma_{\nu_1}\,\gamma_{\nu_2}\,\slashed{q}\,v_3\,\; ,
\notag\\[.2cm]\,\bar{u}_1\,\gamma^{\nu_1}\,\gamma^{\nu_2}\,\slashed{p}_3\,u_0\,\bar{u}_2\,\slashed{p}_1\,\gamma_{\nu_1}\,\gamma_{\nu_2}\,v_3\,\; ,
\qquad \,\bar{u}_1\,\slashed{p}_2\,\gamma^{\mu}\,\slashed{p}_3\,u_0\,\bar{u}_2\,\gamma_{\mu}\,v_3\,\; ,
\qquad \,\bar{u}_1\,\slashed{p}_2\,\gamma^{\mu}\,\slashed{p}_3\,u_0\,\bar{u}_2\,\slashed{p}_1\,\gamma_{\mu}\,\slashed{q}\,v_3\,\; ,
\notag\\[.2cm]\,\bar{u}_1\,\slashed{p}_2\,\gamma^{\nu_1}\,\gamma^{\nu_2}\,u_0\,\bar{u}_2\,\gamma_{\nu_1}\,\gamma_{\nu_2}\,\slashed{q}\,v_3\,\; ,
\qquad \,\bar{u}_1\,\slashed{p}_2\,\gamma^{\nu_1}\,\gamma^{\nu_2}\,u_0\,\bar{u}_2\,\slashed{p}_1\,\gamma_{\nu_1}\,\gamma_{\nu_2}\,v_3\,\; ,
\qquad \,\bar{u}_1\,\slashed{p}_2\,u_0\,\bar{u}_2\,\slashed{p}_1\,v_3\,\; ,
\notag\\[.2cm]\,\bar{u}_1\,\slashed{p}_3\,u_0\,\bar{u}_2\,\slashed{p}_1\,v_3\,\; .  \label{eq:spin1lqQQ}
\end{gather}
Since the singularities of the one-loop splitting operators given in Eq.~\eqref{eq:Ioperator} have the same spin structure as the tree-level splitting operators, the coefficients of the structures \eqref{eq:spin1lqQQ} are regular in $\epsilon$. Finally, the four-dimensional projection Eq.~\eqref{eq:4Dprojection} transforms the linear combination of \eqref{eq:spin0lqQQ} and \eqref{eq:spin1lqQQ} into a linear combination of just:
\begin{equation}
\,\bar{u}_1\,\slashed{p}_2\,u_0\,\bar{u}_2\,\slashed{p}_1\,v_3\,\; ,
\qquad \,\bar{u}_1\,\slashed{p}_2\,u_0\,\bar{u}_2\,\slashed{q}\,v_3\,\; ,
\qquad \,\bar{u}_1\,\slashed{p}_3\,u_0\,\bar{u}_2\,\slashed{p}_1\,v_3\,\; ,
\qquad \,\bar{u}_1\,\slashed{p}_3\,u_0\,\bar{u}_2\,\slashed{q}\,v_3\,\; .
\end{equation}

\subsubsection*{\underline{$q \to q_1q_2\bar{q}_3$}}

Up to the additional permutation $1 \leftrightarrow 2$, the color and spin structures in this case are very similar to those of the $q \to q_1q'_2\bar{q}'_3$ case.
The occurring color structures are, in fact, the same:
\begin{equation}
    \delta_{c_1c_0}\delta_{c_2c_3} \; , \qquad \delta_{c_2c_0}\delta_{c_1c_3} \; .
\end{equation}
The number of spin structures at tree-level is doubled:
\begin{gather}
\,\bar{u}_1\,\gamma^{\mu}\,u_0\,\bar{u}_2\,\gamma_{\mu}\,v_3\,\; ,
\qquad \,\bar{u}_1\,\slashed{p}_2\,u_0\,\bar{u}_2\,\slashed{q}\,v_3\,\; ,
\qquad \,\bar{u}_1\,\slashed{p}_3\,u_0\,\bar{u}_2\,\slashed{q}\,v_3\,\; ,
\notag\\[.2cm]\,\bar{u}_2\,\gamma^{\mu}\,u_0\,\bar{u}_1\,\gamma_{\mu}\,v_3\,\; ,
\qquad \,\bar{u}_2\,\slashed{p}_1\,u_0\,\bar{u}_1\,\slashed{q}\,v_3\,\; ,
\qquad \,\bar{u}_2\,\slashed{p}_3\,u_0\,\bar{u}_1\,\slashed{q}\,v_3\,\; ,
\end{gather}
while the additional spin structures with finite coefficients at one-loop level are:
\begin{gather}
\,\bar{u}_1\,\gamma^{\mu_1}\,\gamma^{\mu_2}\,\gamma^{\mu_3}\,u_0\,\bar{u}_2\,\gamma_{\mu_1}\,\gamma_{\mu_2}\,\gamma_{\mu_3}\,v_3\,\; ,
\qquad \,\bar{u}_1\,\gamma^{\mu}\,u_0\,\bar{u}_2\,\slashed{p}_1\,\gamma_{\mu}\,\slashed{q}\,v_3\,\; ,
\qquad \,\bar{u}_1\,\gamma^{\nu_1}\,\gamma^{\nu_2}\,\slashed{p}_3\,u_0\,\bar{u}_2\,\gamma_{\nu_1}\,\gamma_{\nu_2}\,\slashed{q}\,v_3\,\; ,
\notag\\[.2cm]\,\bar{u}_1\,\gamma^{\nu_1}\,\gamma^{\nu_2}\,\slashed{p}_3\,u_0\,\bar{u}_2\,\slashed{p}_1\,\gamma_{\nu_1}\,\gamma_{\nu_2}\,v_3\,\; ,
\qquad \,\bar{u}_1\,\slashed{p}_2\,\gamma^{\mu}\,\slashed{p}_3\,u_0\,\bar{u}_2\,\gamma_{\mu}\,v_3\,\; ,
\qquad \,\bar{u}_1\,\slashed{p}_2\,\gamma^{\mu}\,\slashed{p}_3\,u_0\,\bar{u}_2\,\slashed{p}_1\,\gamma_{\mu}\,\slashed{q}\,v_3\,\; ,
\notag\\[.2cm]\,\bar{u}_1\,\slashed{p}_2\,\gamma^{\nu_1}\,\gamma^{\nu_2}\,u_0\,\bar{u}_2\,\gamma_{\nu_1}\,\gamma_{\nu_2}\,\slashed{q}\,v_3\,\; ,
\qquad \,\bar{u}_1\,\slashed{p}_2\,\gamma^{\nu_1}\,\gamma^{\nu_2}\,u_0\,\bar{u}_2\,\slashed{p}_1\,\gamma_{\nu_1}\,\gamma_{\nu_2}\,v_3\,\; ,
\qquad \,\bar{u}_1\,\slashed{p}_2\,u_0\,\bar{u}_2\,\slashed{p}_1\,v_3\,\; ,
\notag\\[.2cm]\,\bar{u}_1\,\slashed{p}_3\,u_0\,\bar{u}_2\,\slashed{p}_1\,v_3\,\; ,
\qquad \,\bar{u}_2\,\gamma^{\mu_1}\,\gamma^{\mu_2}\,\gamma^{\mu_3}\,u_0\,\bar{u}_1\,\gamma_{\mu_1}\,\gamma_{\mu_2}\,\gamma_{\mu_3}\,v_3\,\; ,
\qquad \,\bar{u}_2\,\gamma^{\mu}\,u_0\,\bar{u}_1\,\slashed{p}_2\,\gamma_{\mu}\,\slashed{q}\,v_3\,\; ,
\notag\\[.2cm]\,\bar{u}_2\,\gamma^{\nu_1}\,\gamma^{\nu_2}\,\slashed{p}_3\,u_0\,\bar{u}_1\,\gamma_{\nu_1}\,\gamma_{\nu_2}\,\slashed{q}\,v_3\,\; ,
\qquad \,\bar{u}_2\,\gamma^{\nu_1}\,\gamma^{\nu_2}\,\slashed{p}_3\,u_0\,\bar{u}_1\,\slashed{p}_2\,\gamma_{\nu_1}\,\gamma_{\nu_2}\,v_3\,\; ,
\qquad \,\bar{u}_2\,\slashed{p}_1\,\gamma^{\mu}\,\slashed{p}_3\,u_0\,\bar{u}_1\,\gamma_{\mu}\,v_3\,\; ,
\notag\\[.2cm]\,\bar{u}_2\,\slashed{p}_1\,\gamma^{\mu}\,\slashed{p}_3\,u_0\,\bar{u}_1\,\slashed{p}_2\,\gamma_{\mu}\,\slashed{q}\,v_3\,\; ,
\qquad \,\bar{u}_2\,\slashed{p}_1\,\gamma^{\nu_1}\,\gamma^{\nu_2}\,u_0\,\bar{u}_1\,\gamma_{\nu_1}\,\gamma_{\nu_2}\,\slashed{q}\,v_3\,\; ,
\qquad \,\bar{u}_2\,\slashed{p}_1\,\gamma^{\nu_1}\,\gamma^{\nu_2}\,u_0\,\bar{u}_1\,\slashed{p}_2\,\gamma_{\nu_1}\,\gamma_{\nu_2}\,v_3\,\; ,
\notag\\[.2cm]\,\bar{u}_2\,\slashed{p}_1\,u_0\,\bar{u}_1\,\slashed{p}_2\,v_3\,\; ,
\qquad \,\bar{u}_2\,\slashed{p}_3\,u_0\,\bar{u}_1\,\slashed{p}_2\,v_3\,\; .
\end{gather}
Finally, the result of the four-dimensional projection Eq.~\eqref{eq:4Dprojection} of the spin structures is given by:
\begin{gather}
\,\bar{u}_1\,\slashed{p}_2\,u_0\,\bar{u}_2\,\slashed{p}_1\,v_3\,\; ,
\qquad \,\bar{u}_1\,\slashed{p}_2\,u_0\,\bar{u}_2\,\slashed{q}\,v_3\,\; ,
\qquad \,\bar{u}_1\,\slashed{p}_3\,u_0\,\bar{u}_2\,\slashed{p}_1\,v_3\,\; ,
\qquad \,\bar{u}_1\,\slashed{p}_3\,u_0\,\bar{u}_2\,\slashed{q}\,v_3\,\; ,
\notag\\[.2cm]\,\bar{u}_2\,\slashed{p}_1\,u_0\,\bar{u}_1\,\slashed{p}_2\,v_3\,\; ,
\qquad \,\bar{u}_2\,\slashed{p}_1\,u_0\,\bar{u}_1\,\slashed{q}\,v_3\,\; ,
\qquad \,\bar{u}_2\,\slashed{p}_3\,u_0\,\bar{u}_1\,\slashed{p}_2\,v_3\,\; ,
\qquad \,\bar{u}_2\,\slashed{p}_3\,u_0\,\bar{u}_1\,\slashed{q}\,v_3\,\; . \notag\\[.2cm]
\end{gather}

\subsubsection*{\underline{$q \to q_1g_2g_3$}}

In this case, there are two color structures at tree-level reflecting the symmetry of the splitting operator with respect to the exchange of the two gluons:
\begin{equation}
    \big( T^{c_2} T^{c_3} \big)_{c_1c_0} \; , \qquad
    \big( T^{c_3} T^{c_2} \big)_{c_1c_0} \; .
\end{equation}
The one additional structure at one-loop level is already symmetric:
\begin{equation}
    \delta_{c_1c_0} \Tr(T^{c_2} T^{c_3}) \; .
\end{equation}
The spin-structures at tree-level are as follows:
\begin{gather}
\,\bar{u}_1\,\slashed{\epsilon}^*_2\,\slashed{\epsilon}^*_3\,\slashed{p}_3\,u_0\,\; ,
\qquad \,\bar{u}_1\,\slashed{p}_2\,\slashed{\epsilon}^*_2\,\slashed{\epsilon}^*_3\,u_0\,\; ,
\qquad \,\bar{u}_1\,\slashed{\epsilon}^*_2\,u_0\,(p_1 \cdot \epsilon^*_3)\; ,
\qquad \,\bar{u}_1\,\slashed{\epsilon}^*_2\,u_0\,(p_2 \cdot \epsilon^*_3)\; ,
\notag\\[.2cm]\,\bar{u}_1\,\slashed{\epsilon}^*_3\,u_0\,(p_1 \cdot \epsilon^*_2)\; ,
\qquad \,\bar{u}_1\,\slashed{\epsilon}^*_3\,u_0\,(p_3 \cdot \epsilon^*_2)\; ,
\qquad \,\bar{u}_1\,\slashed{p}_2\,u_0\,(\epsilon^*_2 \cdot \epsilon^*_3)\; ,
\qquad \,\bar{u}_1\,\slashed{p}_3\,u_0\,(\epsilon^*_2 \cdot \epsilon^*_3)\; ,
\end{gather}
while the list is extended at one-loop by:
\begin{gather}
\,\bar{u}_1\,\slashed{p}_2\,\slashed{\epsilon}^*_2\,\slashed{p}_3\,u_0\,(p_1 \cdot \epsilon^*_3)\; ,
\qquad \,\bar{u}_1\,\slashed{p}_2\,\slashed{\epsilon}^*_2\,\slashed{p}_3\,u_0\,(p_2 \cdot \epsilon^*_3)\; ,
\qquad \,\bar{u}_1\,\slashed{p}_2\,\slashed{\epsilon}^*_3\,\slashed{p}_3\,u_0\,(p_1 \cdot \epsilon^*_2)\; ,
\notag\\[.2cm]\,\bar{u}_1\,\slashed{p}_2\,\slashed{\epsilon}^*_3\,\slashed{p}_3\,u_0\,(p_3 \cdot \epsilon^*_2)\; ,
\qquad \,\bar{u}_1\,\slashed{p}_2\,u_0\,(p_1 \cdot \epsilon^*_2)(p_1 \cdot \epsilon^*_3)\; ,
\qquad \,\bar{u}_1\,\slashed{p}_2\,u_0\,(p_1 \cdot \epsilon^*_2)(p_2 \cdot \epsilon^*_3)\; ,
\notag\\[.2cm]\,\bar{u}_1\,\slashed{p}_2\,u_0\,(p_1 \cdot \epsilon^*_3)(p_3 \cdot \epsilon^*_2)\; ,
\qquad \,\bar{u}_1\,\slashed{p}_2\,u_0\,(p_2 \cdot \epsilon^*_3)(p_3 \cdot \epsilon^*_2)\; ,
\qquad \,\bar{u}_1\,\slashed{p}_3\,u_0\,(p_1 \cdot \epsilon^*_2)(p_1 \cdot \epsilon^*_3)\; ,
\notag\\[.2cm]\,\bar{u}_1\,\slashed{p}_3\,u_0\,(p_1 \cdot \epsilon^*_2)(p_2 \cdot \epsilon^*_3)\; ,
\qquad \,\bar{u}_1\,\slashed{p}_3\,u_0\,(p_1 \cdot \epsilon^*_3)(p_3 \cdot \epsilon^*_2)\; ,
\qquad \,\bar{u}_1\,\slashed{p}_3\,u_0\,(p_2 \cdot \epsilon^*_3)(p_3 \cdot \epsilon^*_2)\; .
\end{gather}
In both cases, Bose symmetry is not explicit due to the simplification algorithm. While this admittedly makes the expressions less elegant, it has no influence on actual calculations using the splitting operators. Finally, the four-dimensional projection Eq.~\eqref{eq:4Dprojection} yields:
\begin{gather}
\,\bar{u}_1\,\slashed{p}_2\,u_0\,(p_1 \cdot \epsilon^*_2)(p_1 \cdot \epsilon^*_3)\; ,
\qquad \,\bar{u}_1\,\slashed{p}_2\,u_0\,(p_1 \cdot \epsilon^*_2)(p_2 \cdot \epsilon^*_3)\; ,
\qquad \,\bar{u}_1\,\slashed{p}_2\,u_0\,(p_1 \cdot \epsilon^*_3)(p_3 \cdot \epsilon^*_2)\; ,
\notag\\[.2cm]\,\bar{u}_1\,\slashed{p}_2\,u_0\,(p_2 \cdot \epsilon^*_3)(p_3 \cdot \epsilon^*_2)\; ,
\qquad \,\bar{u}_1\,\slashed{p}_3\,u_0\,(p_1 \cdot \epsilon^*_2)(p_1 \cdot \epsilon^*_3)\; ,
\qquad \,\bar{u}_1\,\slashed{p}_3\,u_0\,(p_1 \cdot \epsilon^*_2)(p_2 \cdot \epsilon^*_3)\; ,
\notag\\[.2cm]\,\bar{u}_1\,\slashed{p}_3\,u_0\,(p_1 \cdot \epsilon^*_3)(p_3 \cdot \epsilon^*_2)\; ,
\qquad \,\bar{u}_1\,\slashed{p}_3\,u_0\,(p_2 \cdot \epsilon^*_3)(p_3 \cdot \epsilon^*_2)\; . \label{eq:4Dspin1lqgg}
\end{gather}

\subsubsection*{\underline{$g \to g_1q_2\bar{q}_3$}}

The external states for this splitting operator are the same as in the previous case. This translates into the same color structures up to index permutation. Hence, at tree level, the color structures are:
\begin{equation}
    \big( T^{c_0} T^{c_1} \big)_{c_3c_2} \; , \qquad
    \big( T^{c_1} T^{c_0} \big)_{c_3c_2} \; ,
\end{equation}
while at one-loop there again appears:
\begin{equation}
    \delta_{c_2c_3} \Tr(T^{c_0} T^{c_1}) \; .
\end{equation}
The spin structures are not related in this trivial way, because the simplification algorithm is not symmetric. We observe, for instance, that there are less spin structures at tree level:
\begin{gather}
\,\bar{u}_2\,\slashed{p}_1\,\slashed{\epsilon}^*_1\,\slashed{\epsilon}_0\,v_3\,\; ,
\qquad \,\bar{u}_2\,\slashed{\epsilon}_0\,v_3\,(p_2 \cdot \epsilon^*_1)\; ,
\qquad \,\bar{u}_2\,\slashed{\epsilon}_0\,v_3\,(p_3 \cdot \epsilon^*_1)\; ,
\qquad \,\bar{u}_2\,\slashed{\epsilon}^*_1\,v_3\,(p_1 \cdot \epsilon_0)\; ,
\notag\\[.2cm]\,\bar{u}_2\,\slashed{p}_1\,v_3\,(\epsilon_0 \cdot \epsilon^*_1)\; ,
\qquad \,\bar{u}_2\,\slashed{q}\,v_3\,(\epsilon_0 \cdot \epsilon^*_1)\; ,
\end{gather}
but more additional spin structures with finite coefficients at one-loop level:
\begin{gather}
\,\bar{u}_2\,\slashed{\epsilon}^*_1\,\slashed{\epsilon}_0\,\slashed{q}\,v_3\,\; ,
\qquad \,\bar{u}_2\,\slashed{\epsilon}^*_1\,v_3\,(p_2 \cdot \epsilon_0)\; ,
\qquad \,\bar{u}_2\,\slashed{p}_1\,\slashed{\epsilon}_0\,\slashed{q}\,v_3\,(p_2 \cdot \epsilon^*_1)\; ,
\qquad \,\bar{u}_2\,\slashed{p}_1\,\slashed{\epsilon}_0\,\slashed{q}\,v_3\,(p_3 \cdot \epsilon^*_1)\; ,
\notag\\[.2cm]\,\bar{u}_2\,\slashed{p}_1\,\slashed{\epsilon}^*_1\,\slashed{q}\,v_3\,(p_1 \cdot \epsilon_0)\; ,
\qquad \,\bar{u}_2\,\slashed{p}_1\,\slashed{\epsilon}^*_1\,\slashed{q}\,v_3\,(p_2 \cdot \epsilon_0)\; ,
\qquad \,\bar{u}_2\,\slashed{p}_1\,v_3\,(p_1 \cdot \epsilon_0)(p_2 \cdot \epsilon^*_1)\; ,
\notag\\[.2cm]\,\bar{u}_2\,\slashed{p}_1\,v_3\,(p_1 \cdot \epsilon_0)(p_3 \cdot \epsilon^*_1)\; ,
\qquad \,\bar{u}_2\,\slashed{p}_1\,v_3\,(p_2 \cdot \epsilon_0)(p_2 \cdot \epsilon^*_1)\; ,
\qquad \,\bar{u}_2\,\slashed{p}_1\,v_3\,(p_2 \cdot \epsilon_0)(p_3 \cdot \epsilon^*_1)\; ,
\notag\\[.2cm]\,\bar{u}_2\,\slashed{q}\,v_3\,(p_1 \cdot \epsilon_0)(p_2 \cdot \epsilon^*_1)\; ,
\qquad \,\bar{u}_2\,\slashed{q}\,v_3\,(p_1 \cdot \epsilon_0)(p_3 \cdot \epsilon^*_1)\; ,
\qquad \,\bar{u}_2\,\slashed{q}\,v_3\,(p_2 \cdot \epsilon_0)(p_2 \cdot \epsilon^*_1)\; ,
\notag\\[.2cm]\,\bar{u}_2\,\slashed{q}\,v_3\,(p_2 \cdot \epsilon_0)(p_3 \cdot \epsilon^*_1)\; .
\end{gather}
A simple relation between the spin structures for $q \to qgg$ and $g \to gq\bar{q}$ is restored after four-dimensional projection, Eq.~\eqref{eq:4Dprojection}, as seen by comparing \eqref{eq:4Dspin1lqgg} with:
\begin{gather}
\,\bar{u}_2\,\slashed{p}_1\,v_3\,(p_1 \cdot \epsilon_0)(p_2 \cdot \epsilon^*_1)\; ,
\qquad \,\bar{u}_2\,\slashed{p}_1\,v_3\,(p_1 \cdot \epsilon_0)(p_3 \cdot \epsilon^*_1)\; ,
\qquad \,\bar{u}_2\,\slashed{p}_1\,v_3\,(p_2 \cdot \epsilon_0)(p_2 \cdot \epsilon^*_1)\; ,
\notag\\[.2cm]\,\bar{u}_2\,\slashed{p}_1\,v_3\,(p_2 \cdot \epsilon_0)(p_3 \cdot \epsilon^*_1)\; ,
\qquad \,\bar{u}_2\,\slashed{q}\,v_3\,(p_1 \cdot \epsilon_0)(p_2 \cdot \epsilon^*_1)\; ,
\qquad \,\bar{u}_2\,\slashed{q}\,v_3\,(p_1 \cdot \epsilon_0)(p_3 \cdot \epsilon^*_1)\; ,
\notag\\[.2cm]\,\bar{u}_2\,\slashed{q}\,v_3\,(p_2 \cdot \epsilon_0)(p_2 \cdot \epsilon^*_1)\; ,
\qquad \,\bar{u}_2\,\slashed{q}\,v_3\,(p_2 \cdot \epsilon_0)(p_3 \cdot \epsilon^*_1)\; .
\end{gather}

\subsubsection*{\underline{$g \to g_1g_2g_3$}}

The color structures of the tree-level pure-gluon splitting operator consist, as for ordinary tree-level gluon-scattering amplitudes, of a trace of a product of fundamental representation generators in all possible color-index permutations:
\begin{gather}
    \Tr(T^{c_0} T^{c_1} T^{c_2} T^{c_3}) \; , \qquad
    \Tr(T^{c_0} T^{c_1} T^{c_3} T^{c_2}) \; , \qquad
    \Tr(T^{c_0} T^{c_2} T^{c_1} T^{c_3}) \; , \notag\\[.2cm]
    \Tr(T^{c_0} T^{c_2} T^{c_3} T^{c_1}) \; , \qquad
    \Tr(T^{c_0} T^{c_3} T^{c_1} T^{c_2}) \; , \qquad
    \Tr(T^{c_0} T^{c_3} T^{c_2} T^{c_1}) \; .
\end{gather}
Disconnected traces appear at one-loop only:
\begin{gather}
    \Tr(T^{c_0} T^{c_1}) \Tr(T^{c_2} T^{c_3}) \; , \qquad
    \Tr(T^{c_0} T^{c_2}) \Tr(T^{c_1} T^{c_3}) \; , \qquad
    \Tr(T^{c_0} T^{c_3}) \Tr(T^{c_1} T^{c_2}) \; .
\end{gather}
The spin structures consist of various contractions of polarisation vectors amongst themselves and with external momenta. With the present algorithm, the list at tree-level reads:
\begin{gather}
(\epsilon_0 \cdot \epsilon^*_1)(\epsilon^*_2 \cdot \epsilon^*_3)\; ,
\qquad (\epsilon_0 \cdot \epsilon^*_2)(\epsilon^*_1 \cdot \epsilon^*_3)\; ,
\qquad (\epsilon_0 \cdot \epsilon^*_3)(\epsilon^*_1 \cdot \epsilon^*_2)\; ,
\notag\\[.2cm](p_1 \cdot \epsilon^*_2)(\epsilon^*_1 \cdot \epsilon^*_3)(p_1 \cdot \epsilon_0)\; ,
\qquad (p_1 \cdot \epsilon^*_3)(\epsilon_0 \cdot \epsilon^*_1)(p_1 \cdot \epsilon^*_2)\; ,
\qquad (p_2 \cdot \epsilon_0)(\epsilon^*_1 \cdot \epsilon^*_2)(p_1 \cdot \epsilon^*_3)\; ,
\notag\\[.2cm](p_2 \cdot \epsilon_0)(\epsilon^*_1 \cdot \epsilon^*_3)(p_1 \cdot \epsilon^*_2)\; ,
\qquad (p_2 \cdot \epsilon^*_1)(\epsilon_0 \cdot \epsilon^*_2)(p_1 \cdot \epsilon^*_3)\; ,
\qquad (p_2 \cdot \epsilon^*_1)(\epsilon^*_2 \cdot \epsilon^*_3)(p_1 \cdot \epsilon_0)\; ,
\notag\\[.2cm](p_2 \cdot \epsilon^*_1)(\epsilon^*_2 \cdot \epsilon^*_3)(p_2 \cdot \epsilon_0)\; ,
\qquad (p_2 \cdot \epsilon^*_3)(\epsilon_0 \cdot \epsilon^*_1)(p_1 \cdot \epsilon^*_2)\; ,
\qquad (p_2 \cdot \epsilon^*_3)(\epsilon_0 \cdot \epsilon^*_2)(p_2 \cdot \epsilon^*_1)\; ,
\notag\\[.2cm](p_2 \cdot \epsilon^*_3)(\epsilon^*_1 \cdot \epsilon^*_2)(p_1 \cdot \epsilon_0)\; ,
\qquad (p_3 \cdot \epsilon^*_1)(\epsilon_0 \cdot \epsilon^*_2)(p_2 \cdot \epsilon^*_3)\; ,
\qquad (p_3 \cdot \epsilon^*_1)(\epsilon_0 \cdot \epsilon^*_3)(p_1 \cdot \epsilon^*_2)\; ,
\notag\\[.2cm](p_3 \cdot \epsilon^*_1)(\epsilon^*_2 \cdot \epsilon^*_3)(p_2 \cdot \epsilon_0)\; ,
\qquad (p_3 \cdot \epsilon^*_2)(\epsilon_0 \cdot \epsilon^*_1)(p_1 \cdot \epsilon^*_3)\; ,
\qquad (p_3 \cdot \epsilon^*_2)(\epsilon_0 \cdot \epsilon^*_3)(p_2 \cdot \epsilon^*_1)\; ,
\notag\\[.2cm](p_3 \cdot \epsilon^*_2)(\epsilon_0 \cdot \epsilon^*_3)(p_3 \cdot \epsilon^*_1)\; ,
\notag\\[.2cm](p_3 \cdot \epsilon^*_2)(\epsilon^*_1 \cdot \epsilon^*_3)(p_1 \cdot \epsilon_0)\; ,
\end{gather}
and it is extended at one-loop by:
\begin{gather}
(p_1 \cdot \epsilon_0)(p_1 \cdot \epsilon^*_2)(p_1 \cdot \epsilon^*_3)(p_2 \cdot \epsilon^*_1)\; ,
\qquad (p_1 \cdot \epsilon_0)(p_1 \cdot \epsilon^*_2)(p_1 \cdot \epsilon^*_3)(p_3 \cdot \epsilon^*_1)\; ,
\notag\\[.2cm](p_1 \cdot \epsilon_0)(p_1 \cdot \epsilon^*_2)(p_2 \cdot \epsilon^*_1)(p_2 \cdot \epsilon^*_3)\; ,
\qquad (p_1 \cdot \epsilon_0)(p_1 \cdot \epsilon^*_2)(p_2 \cdot \epsilon^*_3)(p_3 \cdot \epsilon^*_1)\; ,
\notag\\[.2cm](p_1 \cdot \epsilon_0)(p_1 \cdot \epsilon^*_3)(p_2 \cdot \epsilon^*_1)(p_3 \cdot \epsilon^*_2)\; ,
\qquad (p_1 \cdot \epsilon_0)(p_1 \cdot \epsilon^*_3)(p_3 \cdot \epsilon^*_1)(p_3 \cdot \epsilon^*_2)\; ,
\notag\\[.2cm](p_1 \cdot \epsilon_0)(p_2 \cdot \epsilon^*_1)(p_2 \cdot \epsilon^*_3)(p_3 \cdot \epsilon^*_2)\; ,
\qquad (p_1 \cdot \epsilon_0)(p_2 \cdot \epsilon^*_3)(p_3 \cdot \epsilon^*_1)(p_3 \cdot \epsilon^*_2)\; ,
\notag\\[.2cm](p_1 \cdot \epsilon^*_2)(p_1 \cdot \epsilon^*_3)(p_2 \cdot \epsilon_0)(p_2 \cdot \epsilon^*_1)\; ,
\qquad (p_1 \cdot \epsilon^*_2)(p_1 \cdot \epsilon^*_3)(p_2 \cdot \epsilon_0)(p_3 \cdot \epsilon^*_1)\; ,
\notag\\[.2cm](p_1 \cdot \epsilon^*_2)(p_2 \cdot \epsilon_0)(p_2 \cdot \epsilon^*_1)(p_2 \cdot \epsilon^*_3)\; ,
\qquad (p_1 \cdot \epsilon^*_2)(p_2 \cdot \epsilon_0)(p_2 \cdot \epsilon^*_3)(p_3 \cdot \epsilon^*_1)\; ,
\notag\\[.2cm](p_1 \cdot \epsilon^*_3)(\epsilon^*_1 \cdot \epsilon^*_2)(p_1 \cdot \epsilon_0)\; ,
\qquad (p_1 \cdot \epsilon^*_3)(p_2 \cdot \epsilon_0)(p_2 \cdot \epsilon^*_1)(p_3 \cdot \epsilon^*_2)\; ,
\qquad (p_1 \cdot \epsilon^*_3)(p_2 \cdot \epsilon_0)(p_3 \cdot \epsilon^*_1)(p_3 \cdot \epsilon^*_2)\; ,
\notag\\[.2cm](p_2 \cdot \epsilon_0)(p_2 \cdot \epsilon^*_1)(p_2 \cdot \epsilon^*_3)(p_3 \cdot \epsilon^*_2)\; ,
\qquad (p_2 \cdot \epsilon_0)(p_2 \cdot \epsilon^*_3)(p_3 \cdot \epsilon^*_1)(p_3 \cdot \epsilon^*_2)\; ,
\qquad (p_2 \cdot \epsilon^*_1)(\epsilon_0 \cdot \epsilon^*_3)(p_1 \cdot \epsilon^*_2)\; ,
\notag\\[.2cm](p_2 \cdot \epsilon^*_3)(\epsilon^*_1 \cdot \epsilon^*_2)(p_2 \cdot \epsilon_0)\; ,
\qquad (p_3 \cdot \epsilon^*_1)(\epsilon_0 \cdot \epsilon^*_2)(p_1 \cdot \epsilon^*_3)\; ,
\qquad (p_3 \cdot \epsilon^*_1)(\epsilon^*_2 \cdot \epsilon^*_3)(p_1 \cdot \epsilon_0)\; ,
\notag\\[.2cm](p_3 \cdot \epsilon^*_2)(\epsilon_0 \cdot \epsilon^*_1)(p_2 \cdot \epsilon^*_3)\; ,
\qquad (p_3 \cdot \epsilon^*_2)(\epsilon^*_1 \cdot \epsilon^*_3)(p_2 \cdot \epsilon_0)\; .
\end{gather}
The four-dimensional projection Eq.~\eqref{eq:4Dprojection} removes contractions of polarisation vectors amongst themselves:
\begin{gather}
(p_1 \cdot \epsilon_0)(p_1 \cdot \epsilon^*_2)(p_1 \cdot \epsilon^*_3)(p_2 \cdot \epsilon^*_1)\; ,
\qquad (p_1 \cdot \epsilon_0)(p_1 \cdot \epsilon^*_2)(p_1 \cdot \epsilon^*_3)(p_3 \cdot \epsilon^*_1)\; ,
\notag\\[.2cm](p_1 \cdot \epsilon_0)(p_1 \cdot \epsilon^*_2)(p_2 \cdot \epsilon^*_1)(p_2 \cdot \epsilon^*_3)\; ,
\qquad (p_1 \cdot \epsilon_0)(p_1 \cdot \epsilon^*_2)(p_2 \cdot \epsilon^*_3)(p_3 \cdot \epsilon^*_1)\; ,
\notag\\[.2cm](p_1 \cdot \epsilon_0)(p_1 \cdot \epsilon^*_3)(p_2 \cdot \epsilon^*_1)(p_3 \cdot \epsilon^*_2)\; ,
\qquad (p_1 \cdot \epsilon_0)(p_1 \cdot \epsilon^*_3)(p_3 \cdot \epsilon^*_1)(p_3 \cdot \epsilon^*_2)\; ,
\notag\\[.2cm](p_1 \cdot \epsilon_0)(p_2 \cdot \epsilon^*_1)(p_2 \cdot \epsilon^*_3)(p_3 \cdot \epsilon^*_2)\; ,
\qquad (p_1 \cdot \epsilon_0)(p_2 \cdot \epsilon^*_3)(p_3 \cdot \epsilon^*_1)(p_3 \cdot \epsilon^*_2)\; ,
\notag\\[.2cm](p_1 \cdot \epsilon^*_2)(p_1 \cdot \epsilon^*_3)(p_2 \cdot \epsilon_0)(p_2 \cdot \epsilon^*_1)\; ,
\qquad (p_1 \cdot \epsilon^*_2)(p_1 \cdot \epsilon^*_3)(p_2 \cdot \epsilon_0)(p_3 \cdot \epsilon^*_1)\; ,
\notag\\[.2cm](p_1 \cdot \epsilon^*_2)(p_2 \cdot \epsilon_0)(p_2 \cdot \epsilon^*_1)(p_2 \cdot \epsilon^*_3)\; ,
\qquad (p_1 \cdot \epsilon^*_2)(p_2 \cdot \epsilon_0)(p_2 \cdot \epsilon^*_3)(p_3 \cdot \epsilon^*_1)\; ,
\notag\\[.2cm](p_1 \cdot \epsilon^*_3)(p_2 \cdot \epsilon_0)(p_2 \cdot \epsilon^*_1)(p_3 \cdot \epsilon^*_2)\; ,
\qquad (p_1 \cdot \epsilon^*_3)(p_2 \cdot \epsilon_0)(p_3 \cdot \epsilon^*_1)(p_3 \cdot \epsilon^*_2)\; ,
\notag\\[.2cm](p_2 \cdot \epsilon_0)(p_2 \cdot \epsilon^*_1)(p_2 \cdot \epsilon^*_3)(p_3 \cdot \epsilon^*_2)\; ,
\qquad (p_2 \cdot \epsilon_0)(p_2 \cdot \epsilon^*_3)(p_3 \cdot \epsilon^*_1)(p_3 \cdot \epsilon^*_2)\; .
\end{gather}
In this case, however, the expression for the splitting operator is much larger after projection than before. Hence, we do not provide it in electronic form. The only advantage of the projection is the removal of terms proportional to the six-dimensional integral of Section~\ref{Sec:LCGBox} for six possible permutations of the external momenta. On the other hand, since we know (and have checked) that these integrals disappear in four dimensions, one can directly set them to zero in the expanded expression up to $\order{\epsilon^0}$.

\subsection{Requirements posed by the construction of a subtraction scheme} \label{Sec:Subtraction}

If one could obtain the exact $\epsilon$-dependence of the Feynman integrals occurring in the expressions of the splitting operators derived with the methods of the previous section, then this paragraph would not be required. Unfortunately, as we will see in Section~\ref{Sec:LCGBox}, there is one integral that cannot, at least at present, be obtained in this generality. Hence, we must understand what is actually required of the results for them to be useful in the construction of a subtraction scheme. The conclusions of this discussion depend on the perturbation-theory order at which the said subtraction scheme is to be valid. Here, we restrict ourselves to N3LO.

To make the argument clear, we begin with the simplest possible subtraction scheme, namely a scheme at NLO (more details in the classic Refs.~\cite{Catani:1996vz, Frixione:1995ms}). Here, the relevant factor in the dimensionally-regulated phase-space integration measure for a selected massless parton takes the form:
\begin{equation}
\iint_0^1 \frac{\dd{\eta}}{\eta^{\epsilon}} \frac{\dd{\xi}}{\xi^{-1+2\epsilon}} \; ,
\end{equation}
where $\eta \equiv \tfrac{1}{2} (1- \cos\theta)$ and $\theta$ is the angle between the three-momentum of the chosen parton and that of another massless parton, while $\xi$ is the normalised energy of the chosen parton. The squared tree-level matrix element has singular asymptotics at vanishing $\eta$, if the two partons are either a quark-anti-quark pair of the same flavor, or one of them is a gluon:
\begin{equation} \label{eq:CollinearSingularityNLO}
\ip{M^{(0)}}{M^{(0)}} \equiv \frac{f(\eta)}{\eta} \quad \sim \quad \frac{f(0)}{\eta} \qquad ( \eta \to 0 ) \; .
\end{equation}
We have suppressed the dependence on the remaining variables parameterising the final state momenta that are irrelevant to the problem. There may also be a singularity at vanishing $\xi$ if the energy is that of a gluon, but it is sufficient to restrict to just the collinear singularity in order to demonstrate the main issue. A subtraction scheme allows to evaluate the phase space integral of the squared matrix element with the help of:
\begin{equation} \label{eq:NLOsubtraction}
\begin{split}
\int_0^1 \frac{\dd{\eta}}{\eta^{1+\epsilon}} f(\eta) &= \Big[ \int_0^1 \frac{\dd{\eta}}{\eta^{1+\epsilon}} \big( f(\eta) - f(0) \big) \Big] +  \Big[ f(0) \int_0^1 \frac{\dd{\eta}}{\eta^{1+\epsilon}} \Big] \\[.2cm]
&= \Big[ \int_0^1 \frac{\dd{\eta}}{\eta^{1+\epsilon}} \big( f(\eta) - f(0) \big) \Big] + \Big[ - \frac{1}{\epsilon} f(0) \Big] \; .
\end{split}    
\end{equation}
The integrand of the integral in the first square bracket contains a {\it subtraction term}, $-f(0) / \eta^{1+\epsilon}$ that makes it integrable even after $\epsilon$-expansion, while the integral in the second square bracket is called the {\it integrated subtraction term}.

Suppose now, that we would like to evaluate double-real contributions to a cross section at N3LO. These are cross-section contributions from processes that have two additional massless partons in the final state with respect to the Born-approximation process. Furthermore, the required matrix elements are evaluated at one-loop order. For simplicity, let us assume that the process is unpolarised. With these assumptions, the following expression:
\begin{equation} \label{eq:TripleCollinearSubtraction}
    - \Big( \frac{2}{s_{123}} \Big)^2 \Big[ \ev{\bm{\hat{P}}^{(1)}_{f_1f_2f_3}}{M_{f\dots}^{(0)}} + 2 \Re \mel{M_{f\dots}^{(0)}}{\bm{\hat{P}}^{(0)}_{f_1f_2f_3}}{M_{f\dots}^{(1)}} \Big] \; ,
\end{equation}
provides a subtraction term for triple-collinear singularities. Indeed, according to Eqs.~\eqref{eq:factorisation}, \eqref{eq:SplittingFunctions} and \eqref{eq:SplittingFunctions1l}, it correctly removes these singularities from:
\begin{equation}
    2\Re\ip{M_{f_1f_2f_3\dots}^{(0)}}{M_{f_1f_2f_3\dots}^{(1)}} \; .
\end{equation}
By extension of Eq.~\eqref{eq:NLOsubtraction}, we also need an integrated subtraction term, which contains at least one explicit $1/\epsilon$ pole. Since the integrated subtraction term is evaluated in the collinear limit (in the example: $f(\eta = 0)$), there is no dependence on the transverse direction, and we can use an averaged splitting function. This is, however, secondary. More importantly, due to the presence of a $1/\epsilon$ pole generated by phase-space integration, we need the splitting functions to at least $\order{\epsilon}$, in order for the cross-section contribution to be correct at $\order{\epsilon^0}$.

There is yet another issue to take into account when evaluating integrated subtraction terms. Let us, for a moment, return to the example of one parton splitting into two, but this time at the one-loop level. Whereas the tree-level matrix-element squared is a rational function of scalar products of momenta, and has the asymptotic \eqref{eq:CollinearSingularityNLO}, one-loop splitting functions for a splitting into two partons scale as $s_{12}^{-\epsilon}$ when $s_{12} \to 0$ or, when expressed through $\eta$, as $1/\eta^{\epsilon}$. This must be taken into account when evaluating the one-loop integrated subtraction term with Eq.~\eqref{eq:NLOsubtraction}, otherwise the coefficient of the $1/\epsilon$ pole will be incorrect. In our case of main interest, the subtraction term \eqref{eq:TripleCollinearSubtraction} scales as $s_{123}^{-\epsilon}/s_{123}^2$ on purely dimensional grounds. However, the triple-collinear splitting functions have further singularities themselves. Indeed, let $\theta_{ij}$ be the angle between the momenta of partons $i$ and $j$, and let $E_k$ be the energy of parton $k$, with $i,j,k \in \{1,2,3\}$. Then, in the worst case scenario of a gluon splitting into three gluons, the following limits are singular additionally to the triple-collinear singularity already present: 1) $\theta_{ij} \to 0$ (iterated single-collinear limit); 2) $E_i \to 0$ (single-soft limit); 3) $\theta_{ij} \to 0$, $E_k \to 0$; 4) $E_i, E_j \to 0$ (double-soft limit); 5) $E_i, E_j \to 0$, $E_i/E_j \to 0$ (iterated single-soft limit); 6) $\theta_{ij} \to 0$, $E_k, E_l \to 0$; 7) $\theta_{ij} \to 0$, $E_k, E_l \to 0$, $E_k/E_l \to 0$. Each of these singular configurations requires subtraction and integrated subtraction terms with the correct scaling in the relevant variables. Up to four poles are thus generated: one for the triple-collinear limit, and one for each of the three limits in case 7). Thus, the splitting functions must be known to $\order{\epsilon^4}$. To be more precise, beyond $\order{\epsilon}$ we only need the approximations to the splitting functions valid in the respective limit.

Let us conclude this discussion by taking a first look at the results for the Feynman integrals presented in full detail in Section~\ref{Sec:Masters}. The results that are available with exact dependence on $\epsilon$ consist of terms of the form:
\begin{gather}
    (\dots)^{-\epsilon} \, \frac{1}{\epsilon} \, {}_2F_1(\dots,x) \; , \qquad
    (\dots)^{-\epsilon} \, \frac{1}{\epsilon^2} \, {}_2F_1(\dots,x) \; , \notag\\[.2cm]
    (\dots)^{-\epsilon} \, (\dots)^\epsilon \, \frac{1}{\epsilon^2} \, {}_2F_1(\dots,x) \; , \qquad
    (\dots)^{-\epsilon} \, \frac{1}{\epsilon} \, F_1(\dots,x,y) \; , 
\end{gather}
where $_2F_1$ is the hypergeometric function, see Appendix~\ref{App:2F1}, and $F_1$ is the Appell function, see Appendix~\ref{App:F1}. The arguments $x$ and $y$ of these functions belong to the unit interval, and the functions are regular at $x \to 0$ and $y \to 0$. The dots in the exponential functions represent kinematic-dependent variables providing the scaling in the singular limits discussed above. From the structure follows, that the results for the splitting functions will be of sufficient quality for N3LO for a given limit corresponding to $x \to 0$, if the hypergeometric functions are expanded up to $\order{\epsilon^5}$ for a single pole $1/\epsilon$, or up to $\order{\epsilon^6}$ for a double pole $1/\epsilon^2$ in the coefficient of the $_2F_1$ function. These expansions are provided in Appendix~\ref{App:2F1}. In case a singular limit corresponds to $x \to 1$, the scaling might be changed due to the branch-point singularity of the $_2F_1$ function. In order to capture the correct behaviour, we provide expansions about unit argument in Appendix~\ref{App:2F1}. The situation is more involved for integrals containing the Appell $F_1$ function. In any of the limits containing a collinear singularity, the triple-collinear splitting functions must factorize into a product of splitting functions for one-to-two splittings. These splitting functions may be expressed through hypergeometric functions as worst, see Ref.~\cite{Kosower:1999rx}. Hence, the Appell function should reduce to hypergeometric functions in these cases. We have verified the correctness of this conclusion. A similar simplification should take place for limits with a single-soft singularity. The latter has an even simpler functional form, see Ref.~\cite{Catani:2000pi}. In the remaining, double-soft-singular case, the Appell $F_1$ functions present in the integrals reduce to hypergeometric functions as well, see Eqs.~\eqref{eq:LCGOffShellTriangle2}, \eqref{eq:LCGBoxTriangle} and \eqref{eq:F1asy}. In the end, the $F_1$ function with arbitrary arguments is only needed to $\order{\epsilon}$. This expansion is given in Eq.~\eqref{eq:F1exp}.

There still remains the highly-nontrivial six-dimensional integral of Section~\ref{Sec:LCGBox}. We have verified that it only contributes to the pure triple-collinear case, as well as to the triple-collinear/double-soft case. In the latter, one could have expected a non-trivial integral with a square root of a Gram determinant, just as we have obtained, in view of the general results for the one-loop double-soft limit presented in Ref.~\cite{Zhu:2020ftr}\footnote{The results of Ref.~\cite{Zhu:2020ftr} are restricted to expansions up to $\order{\epsilon^0}$. Nevertheless, the discussion of the calculation of the integrals points to the presence of a square root of a Gram determinant at higher orders in the $\epsilon$-expansion}.

\subsection{A short list of checks}

The complexity of the calculation warrants extensive checking. We have successfully performed the following checks:
\begin{enumerate}
    \item comparison of the predicted singularity structure of the splitting operators, Eq.~\eqref{eq:Ioperator}, with that obtained from the direct calculation;
    \item comparison of the anti-symmetric part of the splitting function for $q \to qq'\bar{q}'$ with the result given in Ref.~\cite{Catani:2003vu};
    \item comparison of the splitting functions for $q \to qq'\bar{q}'$ and $q \to qq\bar{q}$ expanded to $\order{\epsilon^0}$ obtained using the methods of Section~\ref{Sec:Diagrams}, with the result of an expansion in the triple-collinear limit of one-loop matrix-elements squared for the processes $V \to q\bar{q}q'\bar{q}'$ and  $V \to q\bar{q}q\bar{q}$, with $V$ a massive off-shell vector boson;
    \item numerical comparison of the triple-collinear limits of one-loop matrix-elements squared at $\order{\epsilon^0}$ for six-parton processes with the predicted asymptotics, Eq.~\eqref{eq:TripleCollinearSubtraction}, using the software library \textsc{NJet} \cite{Badger:2012pg};
    \item comparison of the values of the master integrals obtained from analytic formulae and from Mellin-Barnes representations up to the provided orders of $\epsilon$-expansion.
\end{enumerate}

\section{Master integrals} \label{Sec:Masters}

\begin{figure}[t]
    \centering
    \includegraphics[width=\textwidth]{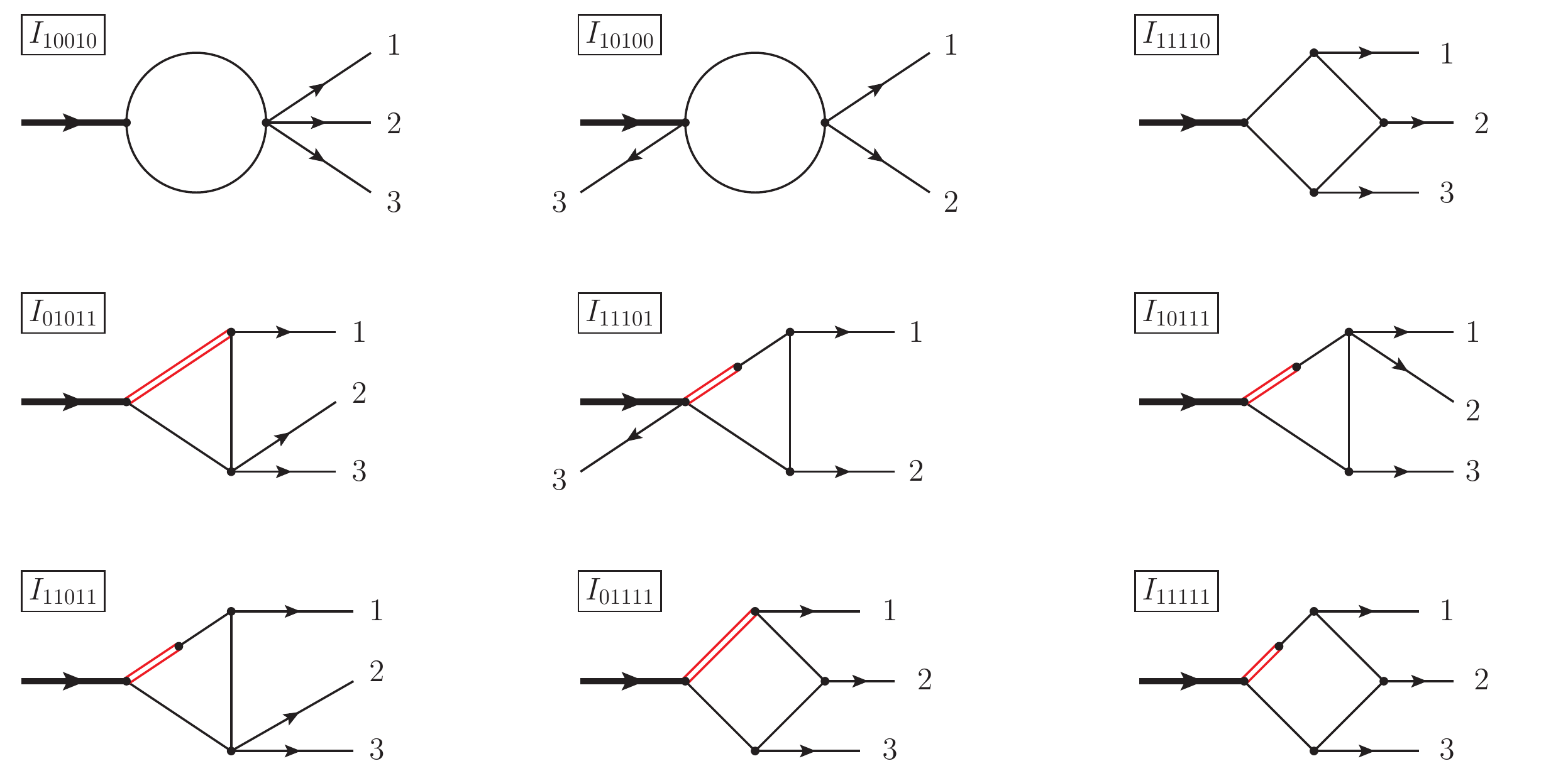}
    \caption{Diagrammatic representation of the master integrals. Lines 1,2 and 3 correspond to outgoing momenta $p_1$, $p_2$ and $p_3$ respectively. The thick incoming line is off shell with momentum $p_{123} = p_1+p_2+p_3$. The red double-line corresponds to a linear (eikonal) propagator introduced through the use of the lightcone gauge.}
    \label{fig:Masters}
\end{figure}

After Passarino-Veltman reduction and integration-by-parts reduction using the software package \textsc{Kira} \cite{Maierhofer:2017gsa, Klappert:2020nbg}, the splitting operators are expressed in terms of 34 master integrals. However, most of these integrals are related by permutation of the external momenta. Taking this into account, there remain only 9 master integrals depicted in Fig.~\ref{fig:Masters}. All of them are generated from the following formula:
\begin{equation} \label{eq:MasterDef}
    \begin{split}
        &I^{(d)}_{a_1 a_2 a_3 a_4 a_5} \equiv \mu^{2\epsilon} \int \frac{\dd[d]{l}}{i \pi^{d/2}} \frac{1}{\big(l^2\big)^{a_1}\big((l+p_1)^2\big)^{a_2}\big((l+p_1+p_2)^2\big)^{a_3}\big((l+p_1+p_2+p_3)^2\big)^{a_4}\big(l \cdot q\big)^{a_5}} \\[.4cm]
        &\quad= (-1)^{a+a_5} \Gamma\Big(a-\frac{d}{2}\Big) \int_{\mathbb{R}_+^4} \Big( \prod_{i=1}^4 \frac{\alpha_i^{a_i-1}\dd{\alpha_i}}{\Gamma(a_i)} \Big) \frac{\delta(1-\sum_{i=1}^4 \alpha_i) (\alpha_1+\alpha_2+\alpha_3+\alpha_4)^{a+a_5-d}}{\big( -s_{12} \; \alpha_1\alpha_3 - s_{123} \; \alpha_1\alpha_4 - s_{23} \; \alpha_2\alpha_4 - i0^+ \big)^{a-d/2}} \\[.2cm]&\quad \qquad\qquad\qquad\qquad\qquad\qquad\qquad\qquad\qquad\times \frac{1}{\big( (p_1 \cdot q) \; \alpha_2 + (p_{12} \cdot q) \; \alpha_3 + (p_{123} \cdot q) \; \alpha_4 - i0^+ \big)^{a_5}} \; ,
    \end{split}
\end{equation}
where $a \equiv a_1+a_2+a_3+a_4$, $p_{ij} \equiv p_i+p_j$, $s_{ij} \equiv p_{ij}^2 = 2 ( p_i \cdot p_j )$, $p_{123} \equiv p_1+p_2+p_3$ and $s_{123} \equiv p_{123}^2 = s_{12}+s_{13}+s_{23}$. Expressions for missing propagators, i.e.\ at $a_i \to 0$ are obtained by the simple replacement:
\begin{equation}
    \frac{\alpha_i^{a_i-1}}{\Gamma(a_i)} \quad \xrightarrow[a_i \; \to \; 0]{} \quad \delta(\alpha_i) \; .
\end{equation}
The results presented in the next subsections are normalised with a universal function of $\epsilon$:
\begin{equation} \label{eq:rGamma}
    r_\Gamma \equiv \frac{\Gamma(1-\epsilon)^2\Gamma(1+\epsilon)}{\Gamma(1-2\epsilon)} \; .
\end{equation}
With $r_\Gamma$, the one-loop triangle integral in $d = 4-2\epsilon$ is a pure double pole in $\epsilon$ up to dependence on the momentum-transfer squared dictated by dimensional arguments:
\begin{equation}
    I^{(4-2\epsilon)}_{11100} = r_\Gamma \; \Big[ \Big(\frac{-s_{12}}{\mu^2}\Big)^{-\epsilon} \frac{1}{s_{12}} \Big] \; \frac{1}{\epsilon^2} \; .
\end{equation}
The master integrals depend on a number of kinematics-based variables. It is useful to define the following dimensionless variables:
\begin{equation}
\begin{gathered}
    x_1 \equiv \frac{s_{23}}{s_{123}} \; , \quad
    x_2 \equiv \frac{s_{13}}{s_{123}} \; , \quad
    x_3 \equiv \frac{s_{12}}{s_{123}} \; , \qquad
    z_i \equiv \frac{p_i \cdot q}{p_{123} \cdot q} \; , \\[.2cm]
    \quad x_i \in (0,1) \; , \qquad z_i \in (0,1) \; , \qquad \sum_{i=1}^3 x_i = \sum_{i=1}^3 z_i = 1 \; . \label{eq:xiziInequalities1}
\end{gathered}
\end{equation}
There are further restrictions on the possible values of $x_i$ and $z_i$. Indeed, consider for example the restframe of $p_{23}$ with $m_{23}^2 \equiv s_{23}$, and let $c_\theta$ be the cosine of the angle between the three-vectors $\bm{p}_1$ and $\bm{q}$. Then:
\begin{equation}
\begin{split}
    x_1 &= \frac{s_{23}}{s_{123}}=\frac{m_{23}^2}{m_{23}^2+2p_1^0 \; m_{23}} = \frac{1}{1+\frac{2p_1^0}{m_{23}}} \; , \\[.4cm]
    1-z_1 &= \frac{p_{23} \cdot q}{p_{123} \cdot q} = \frac{m_{23} \; q^0}{p_1^0 q^0 \left(1-c_{\theta}\right)+ m_{23} \; q^0} = \frac{1}{1+\frac{2p_1^0}{m_{23}}\left(\frac{1-c_{\theta}}{2}\right)} \geq \frac{1}{1+\frac{2p_1^0}{m_{23}}} = x_1 \; .
\end{split}
\end{equation}
Hence, in general:
\begin{equation} \label{eq:xiziInequalities2}
    0 < x_i + z_i \leq 1 \; , \qquad \frac{z_i}{1-x_i} \leq 1 \; , \qquad \frac{x_i}{1-z_i} \leq 1 \; .
\end{equation}

\subsection{Results with exact dependence on the spacetime dimension}

The integrals listed in this subsection have been evaluated in $d = 4-2\epsilon$ dimensions. Their dependence on the dimensional regularisation parameter $\epsilon$ is exact, while the results are expressed in terms of hypergeometric $_2F_1$ functions, see Appendix~\ref{App:2F1}, and Appell $F_1$ functions, see Appendix~\ref{App:F1}. The arguments of these functions belong to the unit interval in every case due to \eqref{eq:xiziInequalities1} and \eqref{eq:xiziInequalities2}. We begin with ordinary Feynman integrals:
\begin{align}
\label{eq:Bubble1}
   I^{(4-2\epsilon)}_{10010} &= r_\Gamma \Big( \frac{-s_{123}}{\mu^2} \Big)^{-\epsilon} \; \frac{1}{\epsilon(1-2\epsilon)} \; , \\[.4cm]
\label{eq:Bubble2}
   I^{(4-2\epsilon)}_{10100} &= r_\Gamma \Big( \frac{-s_{12}}{\mu^2} \Big)^{-\epsilon} \; \frac{1}{\epsilon(1-2\epsilon)} \; , \\[.4cm]
\label{eq:OffShellBox}
   I^{(4-2\epsilon)}_{11110} &= r_\Gamma \Big( \frac{-s_{12}s_{23}}{\mu^2 s_{123}} \Big)^{-\epsilon} \; \frac{2}{s_{12}s_{23}\;\epsilon^2} \notag\\[.2cm]&\quad\times \Big( (x_1+x_2)^\epsilon \; {}_2F_1\Big(-\epsilon,-\epsilon,1-\epsilon,1-\frac{x_1}{x_1+x_2}\Big)+ (x_2+x_3)^\epsilon \; {}_2F_1\Big(-\epsilon,-\epsilon,1-\epsilon,1-\frac{x_3}{x_2+x_3}\Big) \notag\\[.2cm]&\qquad\quad- \big( (x_1+x_2)(x_2+x_3) \big)^\epsilon \; {}_2F_1\Big(-\epsilon,-\epsilon,1-\epsilon,1-\frac{x_1}{x_1+x_2}\frac{x_3}{x_2+x_3}\Big) \Big) \; .
\end{align}
The integrals Eqs.~\eqref{eq:Bubble1} and \eqref{eq:Bubble2} are two cases of the one-loop bubble integral also known as the $B_0$ function. As such, they are textbook material. The one-loop off-shell box integral Eq.~\eqref{eq:OffShellBox} has been evaluated in Ref.~\cite{Bern:1993kr}. The following integrals with a linear propagator can be found in Ref.~\cite{Sborlini:2014eib}:
\begin{align}
\label{eq:LCGTriangleBubble}
   I^{(4-2\epsilon)}_{01011} &= - r_\Gamma \Big( \frac{-s_{23}}{\mu^2} \Big)^{-\epsilon} \; \frac{1}{\big( p_{123}\cdot q \big) \; \epsilon \; (1-2\epsilon)} \; _2F_1\big(1,1-\epsilon,2-2\epsilon,1-z_1\big) \; , \\[.4cm]
\label{eq:LCGTriangle}
    I^{(4-2\epsilon)}_{11101} &= - r_\Gamma \Big( \frac{-s_{12}}{\mu^2} \Big)^{-\epsilon} \; \frac{2}{s_{12} \; \big( p_{12}\cdot q \big) \; \epsilon^2} \; {}_2F_1\Big(1,1,1-\epsilon,1-\frac{z_1}{z_1+z_2}\Big) \; , \\[.4cm]
\label{eq:LCGOffShellTriangle1}
    I^{(4-2\epsilon)}_{10111} &= - r_\Gamma \Big( \frac{-s_{123}}{\mu^2} \Big)^{-\epsilon} \frac{2}{(s_{13}+s_{23}) \; \big( p_{123}\cdot q \big) \; \epsilon \; (1+\epsilon)} \notag\\[.2cm]&\quad\times \Bigg(\,_2F_1\Big(1,1,2+\epsilon,1-\frac{z_3}{1-x_3}\Big)-\frac{x_3^{-\epsilon}}{1-z_3} \; _2F_1\Big(1,1,2+\epsilon,1-\frac{z_3}{1-x_3}\frac{x_3}{1-z_3}\Big)\Bigg) \; .
\end{align}
These results may be verified by directly integrating the corresponding Feynman-parameter integrals using the general expression Eq.~\eqref{eq:MasterDef}. In the present work, we have obtained two more exact results:
\begin{align}
\label{eq:LCGOffShellTriangle2}
    I^{(4-2\epsilon)}_{11011} &= - r_\Gamma \Big( \frac{-s_{123}}{\mu^2} \Big)^{-\epsilon} \frac{1}{(s_{12}+s_{13}) \; \big( p_{123}\cdot q \big) \; \epsilon^2} \notag\\[.2cm]&\quad\times \Bigg(2\,_2F_1\Big(1,1,1-\epsilon,1-\frac{z_1}{1-x_1}\Big)-x_1^{-\epsilon}F_1\Big(1,-\epsilon,1,1-2\epsilon,1-z_1,1-\frac{z_1}{1-x_1}\Big)\Bigg) \; , \\[.4cm]
\label{eq:LCGBoxTriangle}
    I^{(4-2\epsilon)}_{01111} &= - r_\Gamma \Big( \frac{-s_{23}}{\mu^2} \Big)^{-\epsilon} \frac{1}{s_{23} \; \big( p_{123}\cdot q \big) \; \epsilon^2} \; F_1\Big(1,-\epsilon,1,1-2\epsilon,1-z_1,1-z_1-z_2\Big) \; .
\end{align}
The values of these integrals in an $\epsilon$-expansion to $\order{\epsilon^0}$ can be found in Ref.~\cite{Sborlini:2014eib}. These expansions are, however, insufficient for applications at N3LO. The first integral, Eq.~\eqref{eq:LCGOffShellTriangle2}, can be evaluated starting from the following Mellin-Barnes representation, which can be derived with standard methods described for example in Ref.~\cite{Smirnov:2006ry}:
\begin{equation}
    \begin{split}
        I^{(4-2\epsilon)}_{11011} &= -\Big( \frac{-s_{123}}{\mu^2} \Big)^{-\epsilon} \frac{1}{s_{123} \; \big( p_{123}\cdot q \big) \; \Gamma(-2\epsilon)} \\[.2cm]&\quad\times\frac{1}{(2\pi i)^2} \iint_{\mathcal{C}} \dd{z_1} \dd{z_2} \Gamma\big(1+\epsilon+z_1\big)\Gamma\big(1+z_2\big)\Gamma\big(1+z_1+z_2\big)\Gamma\big(-\epsilon-z_1\big)\Gamma\big(-1-\epsilon-z_2\big) \\[.2cm]&\qquad\qquad\qquad\qquad\qquad\times \Gamma\big(-z_1\big)\Gamma\big(-z_2\big) \; \Big( \frac{s_{23}}{s_{123}} \Big)^{z_1} \Big( \frac{p_1 \cdot q}{p_{123} \cdot q} \Big)^{z_2} \; .
    \end{split}
\end{equation}
Using the method proposed in Appendix C of Ref.~\cite{Anastasiou:2013srw}, one obtains a one-fold integral, which straightforwardly yields Eq.~\eqref{eq:LCGOffShellTriangle2}. The second integral, Eq.~\eqref{eq:LCGBoxTriangle}, has the Mellin-Barnes representation:
\begin{equation}
    \begin{split}
        I^{(4-2\epsilon)}_{01111} &= -\Big( \frac{-s_{23}}{\mu^2} \Big)^{-\epsilon} \frac{\Gamma(1+\epsilon)}{s_{23} \; \big( p_{123}\cdot q \big) \; \Gamma(-2\epsilon)} \\[.2cm]&\quad\times\frac{1}{(2\pi i)^2} \iint_{\mathcal{C}} \dd{z_1} \dd{z_2} \Gamma\big(-\epsilon+z_1\big)\Gamma\big(1+z_2\big)\Gamma\big(1+z_1+z_2\big)\Gamma\big(-1-\epsilon-z_1-z_2\big) \\[.2cm]&\qquad\qquad\qquad\qquad\qquad\times \Gamma\big(-z_1\big)\Gamma\big(-z_2\big) \; \Big( \frac{p_1 \cdot q}{p_{123} \cdot q} \Big)^{z_1} \Big( \frac{p_{12} \cdot q}{p_{123} \cdot q} \Big)^{z_2} \; .
    \end{split}
\end{equation}
The result follows by comparing with the Mellin-Barnes representation of the Appell $F_1$ function that appears last in Eq.~\eqref{eq:F1def}. An analysis of the soft/collinear limits of the splitting functions shows that the function $F_1(1,-\epsilon,1,1-2\epsilon,1-x,1-y)$ with general arguments $x \neq y$ away from the endpoints 0 and 1, is only needed in an $\epsilon$-expansion to $\order{\epsilon}$. The respective result can be found in Eq.~\eqref{eq:F1exp}. Indeed, in any soft/collinear limit, the $F_1$ functions with general arguments cancel, while they are expressible through $_2F_1$ functions in the remaining cases. The appropriate expressions are given in Appendix~\ref{App:F1}.

\subsection{The light-cone-gauge box integral with one leg off shell} \label{Sec:LCGBox}

It turns out that the light-cone-gauge box integral $I^{(d)}_{11111}$ is convergent in $d=6$ dimensions. This property mostly follows from the well-known fact that there are no soft/collinear divergences in six dimensions, and from ultraviolet power counting. In the case of ordinary Feynman integrals, these arguments are sufficient to prove convergence. Due to the presence of the linear propagator, it is necessary to consider possible rapidity divergences. The latter make the six-dimensional integrals $I^{(6)}_{01111}, I^{(6)}_{10111},I^{(6)}_{11011}$ and $I^{(6)}_{11101}$ diverge, although the aforementioned ``soft/collinear/ultraviolet argument''
 applies in their case as well. Since these rapidity divergences are due to large longitudinal, $l_+$, and transverse, $\bm{l}_T$, components of the loop momentum defined with respect to $q$, with $|l_+| \sim \bm{l}_T^2$, they only disappear for integrals with at least four ordinary propagators, e.g.\ for $I^{(6)}_{11111}$. With the help of the integration-by-parts reduction relations and using the methods of Refs.~\cite{Tarasov:1996br, Tarasov:1997kx}, we obtain the following dimensional-shift relation:
\begin{align} \label{eq:DimShift}
    2 s_{123} \; {I}^{{(d)}}_{11111} &= \frac{\big(x_1z_1-x_2z_2+x_3z_3\big)^2-4x_1x_3z_1z_3}{x_1x_3z_1\big(1-x_3-z_3\big)} \; \big(d-4\big) \; {I}^{{(d+2)}}_{11111} \notag\\[.4cm]
    &\quad+\frac{\big(x_1z_1-x_2z_2+x_3z_3\big)\big(z_1+z_2\big)-2x_3z_1z_3}{x_3z_1\big(1-x_3-z_3\big)} \; {I}^{{(d)}}_{01111}-\frac{x_1z_1-x_2z_2+x_3z_3}{x_1x_3z_1} \; {I}^{{(d)}}_{10111}\notag\\[.2cm]&\quad+\frac{\big(x_1z_1-x_2z_2+x_3z_3\big)-2x_1x_3}{x_1x_3\big(1-x_3-z_3\big)} \; {I}^{{(d)}}_{11011}-\frac{\big(x_1z_1-x_2z_2+x_3z_3\big)-2x_1(z_1+z_2)}{x_1\big(1-x_3-z_3\big)} \; {I}^{{(d)}}_{11101}\notag\\[.2cm]&\quad+\frac{\big(x_1z_1-x_2z_2+x_3z_3\big)-2z_1(x_1+x_2)}{z_1\big(1-x_3-z_3\big)} \; \Big( \frac{s_{123}}{p_{123} \cdot q} \Big) \; {I}^{{(d)}}_{11110} \; .
\end{align}
The presence of the factor $(d-4)$ on the right-hand-side coefficient of $I^{(d+2)}_{11111}$ implies that the integral $I^{(4-2\epsilon)}_{11111}$ expanded up to $\order{\epsilon^0}$ is entirely given by the integrals of the previous subsection. In fact, this expansion is given in Ref.~\cite{Sborlini:2014eib}. On the other hand, since the results of previous studies of one-loop triple-collinear splittings have been restricted to expansions up to $\order{\epsilon^0}$, this integral has never been really needed in view of Eq.~\eqref{eq:DimShift}. Let us note that the coefficient of $I^{(d+2)}_{11111}$ is proportional to the Gram determinant of the momenta $k_i=p_i$, $i=1,2,3$, $k_4 = q$. Indeed:
\begin{equation} \label{eq:GramDet}
\begin{split}
    \Delta_4 \equiv \det(k_i \cdot k_j) &= \frac{1}{4} s^2_{123} \, ( p_{123} \cdot q )^2 \big( \big(x_1z_1-x_2z_2+x_3z_3\big)^2-4x_1x_3z_1z_3 \big) \\[.4cm]
    &\equiv \Big( \frac{1}{2} s_{123} \, ( p_{123} \cdot q ) \; d_4 \Big)^2 \; .
\end{split}
\end{equation}
An analysis of the soft/collinear limits of the splitting functions shows that only the value of $I^{(6)}_{11111}$ is required for non-singular configurations, while the double-soft limit $p_1,p_2 \to 0$, with $z_1/(z_1+z_2) \in (0,1)$, i.e.\ excluding the strongly ordered limits $z_1 \ll z_2$ and $z_2 \ll z_1$ further requires $I^{(6-2\epsilon)}_{11111}$ up to $\order{\epsilon}$. In order to evaluate $I^{(6-2\epsilon)}_{11111}$, we turn to the Feynman-parameter representation Eq.~\eqref{eq:MasterDef}:
\begin{equation} \label{eq:I11111}
    \begin{split}
        I^{(6-2\epsilon)}_{11111} &=  \frac{\Big( \frac{-s_{123}}{\mu^2} \Big)^{-\epsilon}\Gamma(1+\epsilon)}{s_{123} \, \big( p_{123} \cdot q \big)} \int_{\mathbb{R}_+^4} \prod_{i=1}^4 \dd{\alpha_i} \frac{\delta(1-\alpha_1) \big(\alpha_1+\alpha_2+\alpha_3+\alpha_4\big)^{2\epsilon-1}}{\big(\alpha_2\,z_1+\alpha_3(z_1+z_2)+\alpha_4\big)\big(\alpha_1(\alpha_3\,x_3+\alpha_4)+\alpha_2\alpha_4\,x_1\big)^{1+\epsilon}} \\[.4cm]
        &= \frac{\Big( \frac{-s_{123}\,( p_1 \cdot q )}{\mu^2\,( p_{123} \cdot q )} \Big)^{-\epsilon}\Gamma(1+\epsilon)}{s_{123} \, \big( p_{12} \cdot q \big)} \iiint_0^\infty \dd{\alpha_2} \dd{\alpha_3} \dd{\alpha_4} \frac{\big(1+\alpha_2+\alpha_3\,y_1+\alpha_4\, z_1\big){}^{2\epsilon-1}}{\big(\alpha_2+\alpha_3+\alpha_4\big)\big(\alpha_3\,u_3+\alpha_4 (1+\alpha_2\, x_1)\big)^{1+\epsilon}} \; .
    \end{split}
\end{equation}
In the first line, we have used the Cheng-Wu theorem to change the $\delta$-function to only restrict $\alpha_1$. In the second line, we have rescaled the integration variables as follows:
\begin{equation} \label{eq:rescaling}
    \alpha_3 \; \to \; \alpha_3 \, y_1 \; , \qquad
    \alpha_4 \; \to \; \alpha_4 \, z_1 \; ,
\end{equation}
and defined:
\begin{equation}
    y_1 \equiv \frac{z_1}{z_1+z_2} \in (0,1) \; , \qquad u_3 \equiv \frac{x_3}{1-z_3} \in (0,1) \; .
\end{equation}
The integral thus depends on the variables $x_1$, $y_1$, $z_1$ and $u_3$. The purpose of the rescaling Eq.~\eqref{eq:rescaling} is to yield an integrable integrand in the double-soft limit. In this limit, out of the four variables $x_1,y_1,z_1,u_3$ only $z_1$ vanishes. Setting $p_1 = p_2 = 0$ in the first line of Eq.~\eqref{eq:I11111} yields an integral in $\alpha_4$ that is not integrable at the lower limit $\alpha_4 = 0$. The integral on the second line does not suffer from this pathological behaviour. At the same time, we factor-out the non-trivial scaling in the limit in the pre-factor.

The integrations may now be performed with the software package \textsc{PolyLogTools} \cite{Duhr:2019tlz} in the order $\alpha_3$, $\alpha_2$, $\alpha_4$, using a fibration basis for generalized polylogarithms corresponding to the ordering of the variables ($\alpha_3$, $\alpha_2$, $\alpha_4$, $z_1$, $x_1$, $y_1$, $u_3$, $r_1$, $r_2$). The last two variables are roots of a quadratic polynomial in $\alpha_4$ that must be factorized in order to obtain an alphabet linear in the integration variables:
\begin{equation}
    r_{1,2} \equiv \frac{x_1z_1-x_2z_2+x_3z_3 \mp d_4}{2x_1z_1z_3} \; .
\end{equation}
Here, $d_4$ is related to the Gram determinant as defined in Eq.~\eqref{eq:GramDet}. The final result is:
\begin{align}
&2\sqrt{\Delta_4} \; \Bigg( \Big( \frac{-s_{123}\,( p_1 \cdot q )}{\mu^2\,( p_{123} \cdot q )} \Big)^{-\epsilon}\Gamma(1+\epsilon) \Bigg)^{-1} \; I^{(6-2\epsilon)}_{11111} = \notag\\[.2cm]&\quad G\left(0,-r_1\right)G\left(0,u_3\right)G\left(\frac{u_3}{r_1},x_1\right)+G\left(0,-r_1\right)G\left(0,u_3\right)G\left(\frac{u_3}{r_1},y_1\right)
\notag\\[.2cm]&-G\left(0,-r_2\right)G\left(0,u_3\right)G\left(\frac{u_3}{r_2},x_1\right)-G\left(0,-r_2\right)G\left(0,u_3\right)G\left(\frac{u_3}{r_2},y_1\right)
\notag\\[.2cm]&+G\left(0,-r_2\right)G\left(0,x_1\right)G\left(\frac{1-x_1}{r_2x_1},z_1\right)-G\left(0,-r_1\right)G\left(0,x_1\right)G\left(\frac{1-x_1}{r_1x_1},z_1\right)
\notag\\[.2cm]&+G\left(0,-r_1\right)G\left(0,y_1\right)G\left(y_1-\frac{1}{r_1},z_1\right)-G\left(0,-r_2\right)G\left(0,y_1\right)G\left(y_1-\frac{1}{r_2},z_1\right)
\notag\\[.2cm]&+G\left(0,-r_1\right)G\left(0,y_1\right)G\left(\frac{u_3}{r_1y_1},x_1\right)-G\left(0,-r_1\right)G\left(0,u_3\right)G\left(\frac{u_3}{r_1y_1},x_1\right)
\notag\\[.2cm]&+G\left(0,-r_2\right)G\left(0,u_3\right)G\left(\frac{u_3}{r_2y_1},x_1\right)-G\left(0,u_3\right)G\left(0,y_1\right)G\left(\frac{u_3}{r_1y_1},x_1\right)
\notag\\[.2cm]&+G\left(0,u_3\right)G\left(0,y_1\right)G\left(\frac{u_3}{r_2y_1},x_1\right)-G\left(0,-r_2\right)G\left(0,y_1\right)G\left(\frac{u_3}{r_2y_1},x_1\right)
\notag\\[.2cm]&+G\left(0,-r_1\right)G\left(0,u_3\right)G\left(\frac{y_1}{u_3}-\frac{1}{r_1},z_1\right)-G\left(0,-r_1\right)G\left(0,y_1\right)G\left(\frac{y_1}{u_3}-\frac{1}{r_1},z_1\right)
\notag\\[.2cm]&+G\left(0,u_3\right)G\left(0,y_1\right)G\left(\frac{y_1}{u_3}-\frac{1}{r_1},z_1\right)-G\left(0,-r_2\right)G\left(0,u_3\right)G\left(\frac{y_1}{u_3}-\frac{1}{r_2},z_1\right)
\notag\\[.2cm]&+G\left(0,-r_2\right)G\left(0,y_1\right)G\left(\frac{y_1}{u_3}-\frac{1}{r_2},z_1\right)-G\left(0,u_3\right)G\left(0,y_1\right)G\left(\frac{y_1}{u_3}-\frac{1}{r_2},z_1\right)
\notag\\[.2cm]&-G\left(0,0,-r_1\right)G\left(0,u_3\right)+G\left(0,0,-r_1\right)G\left(\frac{1}{r_1},x_1\right)-G\left(0,0,-r_1\right)G\left(1-\frac{1}{r_1},z_1\right)
\notag\\[.2cm]&-G\left(0,0,-r_1\right)G\left(\frac{u_3}{r_1},x_1\right)-G\left(0,0,-r_1\right)G\left(\frac{u_3}{r_1},y_1\right)+G\left(0,0,-r_1\right)G\left(\frac{1}{r_1},y_1\right)
\notag\\[.2cm]&+G\left(0,0,-r_1\right)G\left(\frac{u_3}{r_1y_1},x_1\right)-G\left(0,0,-r_1\right)G\left(\frac{y_1}{u_3}-\frac{1}{r_1},z_1\right)+G\left(0,0,-r_1\right)G\left(y_1-\frac{1}{r_1},z_1\right)
\notag\\[.2cm]&+G\left(0,0,-r_2\right)G\left(0,u_3\right)-G\left(0,0,-r_2\right)G\left(\frac{1}{r_2},x_1\right)+G\left(0,0,-r_2\right)G\left(1-\frac{1}{r_2},z_1\right)
\notag\\[.2cm]&+G\left(0,0,-r_2\right)G\left(\frac{u_3}{r_2},x_1\right)+G\left(0,0,-r_2\right)G\left(\frac{u_3}{r_2},y_1\right)-G\left(0,0,-r_2\right)G\left(\frac{1}{r_2},y_1\right)
\notag\\[.2cm]&-G\left(0,0,-r_2\right)G\left(\frac{u_3}{r_2y_1},x_1\right)+G\left(0,0,-r_2\right)G\left(\frac{y_1}{u_3}-\frac{1}{r_2},z_1\right)-G\left(0,0,-r_2\right)G\left(y_1-\frac{1}{r_2},z_1\right)
\notag\\[.2cm]&-G\left(0,0,u_3\right)G\left(\frac{u_3}{r_1},x_1\right)+G\left(0,-r_1\right)G\left(0,0,u_3\right)-G\left(0,-r_2\right)G\left(0,0,u_3\right)
\notag\\[.2cm]&+G\left(0,0,u_3\right)G\left(\frac{u_3}{r_2},x_1\right)-G\left(0,0,u_3\right)G\left(\frac{u_3}{r_1},y_1\right)+G\left(0,0,u_3\right)G\left(\frac{u_3}{r_2},y_1\right)
\notag\\[.2cm]&+G\left(0,0,u_3\right)G\left(\frac{u_3}{r_1y_1},x_1\right)-G\left(0,0,u_3\right)G\left(\frac{u_3}{r_2y_1},x_1\right)-G\left(0,0,u_3\right)G\left(\frac{y_1}{u_3}-\frac{1}{r_1},z_1\right)
\notag\\[.2cm]&+G\left(0,0,u_3\right)G\left(\frac{y_1}{u_3}-\frac{1}{r_2},z_1\right)-G\left(0,0,x_1\right)G\left(\frac{1-x_1}{r_1x_1},z_1\right)+G\left(0,0,x_1\right)G\left(\frac{1-x_1}{r_2x_1},z_1\right)
\notag\\[.2cm]&+G\left(0,0,y_1\right)G\left(\frac{u_3}{r_1y_1},x_1\right)+G\left(0,0,y_1\right)G\left(y_1-\frac{1}{r_1},z_1\right)-G\left(0,0,y_1\right)G\left(y_1-\frac{1}{r_2},z_1\right)
\notag\\[.2cm]&-G\left(0,0,y_1\right)G\left(\frac{u_3}{r_2y_1},x_1\right)-G\left(0,0,y_1\right)G\left(\frac{y_1}{u_3}-\frac{1}{r_1},z_1\right)+G\left(0,0,y_1\right)G\left(\frac{y_1}{u_3}-\frac{1}{r_2},z_1\right)
\notag\\[.2cm]&-G\left(0,-r_1\right)G\left(1,0,u_3\right)+G\left(0,-r_2\right)G\left(1,0,u_3\right)-G\left(0,-r_1\right)G\left(1,0,x_1\right)
\notag\\[.2cm]&+G\left(1,0,x_1\right)G\left(\frac{1-x_1}{r_1x_1},z_1\right)-G\left(1,0,x_1\right)G\left(\frac{1-x_1}{r_2x_1},z_1\right)+G\left(0,-r_2\right)G\left(1,0,x_1\right)
\notag\\[.2cm]&-G\left(0,-r_1\right)G\left(1,0,y_1\right)+G\left(0,-r_2\right)G\left(1,0,y_1\right)+G\left(0,-r_1\right)G\left(1-\frac{1}{r_1},-\frac{1}{r_1},z_1\right)
\notag\\[.2cm]&+G\left(0,-r_1\right)G\left(\frac{1}{r_1},0,x_1\right)+G\left(0,-r_1\right)G\left(\frac{1}{r_1},0,y_1\right)-G\left(0,-r_2\right)G\left(1-\frac{1}{r_2},-\frac{1}{r_2},z_1\right)
\notag\\[.2cm]&-G\left(0,-r_1\right)G\left(\frac{u_3}{r_1},0,x_1\right)-G\left(0,-r_2\right)G\left(\frac{1}{r_2},0,x_1\right)-G\left(0,-r_2\right)G\left(\frac{1}{r_2},0,y_1\right)
\notag\\[.2cm]&+G\left(0,u_3\right)G\left(\frac{u_3}{r_1},0,x_1\right)-G\left(0,-r_1\right)G\left(\frac{u_3}{r_1},0,y_1\right)+G\left(0,u_3\right)G\left(\frac{u_3}{r_1},0,y_1\right)
\notag\\[.2cm]&+G\left(0,-r_2\right)G\left(\frac{u_3}{r_2},0,x_1\right)-G\left(0,u_3\right)G\left(\frac{u_3}{r_2},0,x_1\right)+G\left(0,-r_2\right)G\left(\frac{u_3}{r_2},0,y_1\right)
\notag\\[.2cm]&-G\left(0,u_3\right)G\left(\frac{u_3}{r_2},0,y_1\right)-G\left(0,x_1\right)G\left(\frac{1-x_1}{r_1x_1},0,z_1\right)-G\left(0,-r_1\right)G\left(\frac{1-x_1}{r_1x_1},-\frac{1}{r_1},z_1\right)
\notag\\[.2cm]&+G\left(0,x_1\right)G\left(\frac{1-x_1}{r_2x_1},0,z_1\right)+G\left(0,-r_2\right)G\left(\frac{1-x_1}{r_2x_1},-\frac{1}{r_2},z_1\right)-G\left(0,-r_1\right)G\left(y_1-\frac{1}{r_1},-\frac{1}{r_1},z_1\right)
\notag\\[.2cm]&+G\left(0,y_1\right)G\left(y_1-\frac{1}{r_1},y_1,z_1\right)+G\left(0,-r_2\right)G\left(y_1-\frac{1}{r_2},-\frac{1}{r_2},z_1\right)-G\left(0,y_1\right)G\left(y_1-\frac{1}{r_2},y_1,z_1\right)
\notag\\[.2cm]&-G\left(0,u_3\right)G\left(\frac{u_3}{r_1y_1},0,x_1\right)+G\left(0,y_1\right)G\left(\frac{u_3}{r_1y_1},0,x_1\right)+G\left(0,-r_1\right)G\left(\frac{u_3}{r_1y_1},0,x_1\right)
\notag\\[.2cm]&+G\left(0,u_3\right)G\left(\frac{u_3}{r_2y_1},0,x_1\right)-G\left(0,y_1\right)G\left(\frac{u_3}{r_2y_1},0,x_1\right)-G\left(0,-r_2\right)G\left(\frac{u_3}{r_2y_1},0,x_1\right)
\notag\\[.2cm]&+G\left(0,u_3\right)G\left(\frac{y_1}{u_3}-\frac{1}{r_1},\frac{y_1}{u_3},z_1\right)-G\left(0,y_1\right)G\left(\frac{y_1}{u_3}-\frac{1}{r_1},\frac{y_1}{u_3},z_1\right)+G\left(0,-r_1\right)G\left(\frac{y_1}{u_3}-\frac{1}{r_1},-\frac{1}{r_1},z_1\right)
\notag\\[.2cm]&-G\left(0,u_3\right)G\left(\frac{y_1}{u_3}-\frac{1}{r_2},\frac{y_1}{u_3},z_1\right)+G\left(0,y_1\right)G\left(\frac{y_1}{u_3}-\frac{1}{r_2},\frac{y_1}{u_3},z_1\right)-G\left(0,-r_2\right)G\left(\frac{y_1}{u_3}-\frac{1}{r_2},-\frac{1}{r_2},z_1\right)
\notag\\[.2cm]&-G\left(-1,0,0,-r_1\right)+G\left(-1,0,0,-r_2\right)+G\left(0,0,0,-r_1\right)-G\left(0,0,0,-r_2\right)
+G\left(1-\frac{1}{r_1},1,0,z_1\right)\notag\\[.2cm]&+G\left(1-\frac{1}{r_1},-\frac{1}{r_1},0,z_1\right)
+G\left(\frac{1}{r_1},0,0,x_1\right)+G\left(\frac{1}{r_1},0,0,y_1\right)-G\left(1-\frac{1}{r_2},1,0,z_1\right)
\notag\\[.2cm]&-G\left(\frac{1}{r_2},0,0,x_1\right)-G\left(\frac{1}{r_2},0,0,y_1\right)-G\left(1-\frac{1}{r_2},-\frac{1}{r_2},0,z_1\right)
-G\left(\frac{u_3}{r_1},0,0,x_1\right)+G\left(\frac{u_3}{r_2},0,0,x_1\right)\notag\\[.2cm]&-G\left(\frac{u_3}{r_1},0,0,y_1\right)
+G\left(\frac{u_3}{r_2},0,0,y_1\right)-G\left(\frac{1-x_1}{r_1x_1},-\frac{1}{r_1},0,z_1\right)+G\left(\frac{1-x_1}{r_2x_1},-\frac{1}{r_2},0,z_1\right)
\notag\\[.2cm]&-G\left(y_1-\frac{1}{r_1},-\frac{1}{r_1},0,z_1\right)-G\left(y_1-\frac{1}{r_1},y_1,0,z_1\right)+G\left(y_1-\frac{1}{r_2},-\frac{1}{r_2},0,z_1\right)
+G\left(\frac{u_3}{r_1y_1},0,0,x_1\right)\notag\\[.2cm]&-G\left(\frac{u_3}{r_2y_1},0,0,x_1\right)+G\left(y_1-\frac{1}{r_2},y_1,0,z_1\right)
+G\left(\frac{y_1}{u_3}-\frac{1}{r_1},-\frac{1}{r_1},0,z_1\right)+G\left(\frac{y_1}{u_3}-\frac{1}{r_1},\frac{y_1}{u_3},0,z_1\right)\notag\\[.2cm]&-G\left(\frac{y_1}{u_3}-\frac{1}{r_2},-\frac{1}{r_2},0,z_1\right)
-G\left(\frac{y_1}{u_3}-\frac{1}{r_2},\frac{y_1}{u_3},0,z_1\right)
+\zeta_2\Bigg(-3G\left(\frac{u_3}{r_1},x_1\right)+3G\left(\frac{u_3}{r_2},x_1\right)\notag\\[.2cm]&-3G\left(\frac{u_3}{r_1},y_1\right)+3G\left(\frac{u_3}{r_2},y_1\right)-G\left(\frac{1-x_1}{r_1x_1},z_1\right)+G\left(\frac{1-x_1}{r_2x_1},z_1\right)+3G\left(\frac{u_3}{r_1y_1},x_1\right)\notag\\[.2cm]&-3G\left(\frac{u_3}{r_2y_1},x_1\right)-3G\left(\frac{y_1}{u_3}-\frac{1}{r_1},z_1\right)+3G\left(\frac{y_1}{u_3}-\frac{1}{r_2},z_1\right)+3G\left(\frac{1}{r_1},x_1\right)-3G\left(\frac{1}{r_2},x_1\right)\notag\\[.2cm]&+3G\left(y_1-\frac{1}{r_1},z_1\right)-3G\left(y_1-\frac{1}{r_2},z_1\right)+3G\left(\frac{1}{r_1},y_1\right)-3G\left(\frac{1}{r_2},y_1\right)-3G\left(1-\frac{1}{r_1},z_1\right)\notag\\[.2cm]&+3G\left(1-\frac{1}{r_2},z_1\right)-3G\left(-1,-r_1\right)+3G\left(-1,-r_2\right)+4G\left(0,-r_1\right)-4G\left(0,-r_2\right)\Bigg) + \order{\epsilon} \; , \label{eq:I11111ep0}
\end{align}
where the multiple polylogarithms are defined recursively as follows:
\begin{equation}
    G(a_1,\dots,a_n,z) \equiv \int_0^z \frac{\dd{t}}{t-a_1} \, G(a_2,\dots,a_n,t) \; , \qquad G(\underbrace{0,\dots,0}_{n},z) \equiv \frac{1}{n!} \ln^n(z) \; ,
\end{equation}
and can be numerically evaluated with \textsc{PolyLogTools} through its interface to the software package \textsc{GiNaC} \cite{Bauer:2000cp, Vollinga:2004sn}. The right-hand-side of Eq.~\eqref{eq:I11111ep0} is regular in $z_1$ by construction. The double-soft limit may be easily obtained with \textsc{PolyLogTools} using the function \texttt{ExpandPolyLogs} to expand in $z_1$ to $\order{z_1^0}$. The result may then be simplified using the shuffle algebra of the multiple polylogarithms with the function \texttt{ShuffleG}. It turns out that the occurring multiple polylogarithms may be rewritten in terms of classical polylogarithms up to weight three only by simply transforming the expression with the function \texttt{GToLi}. Of course, the result Eq.~\eqref{eq:I11111ep0} may also be rewritten in terms of classical polylogarithms up to weight three, since arbitrary multiple polylogarithms up to this weight have this property. Apart from the double-soft limit, however, the result is not simpler, but involves even more polylogarithms than there are multiple polylogarithms in Eq.~\eqref{eq:I11111ep0}. Finally, we relegate the lengthy expression for the $\order{\epsilon}$ term of the expansion of the left-hand-side of Eq.~\eqref{eq:I11111ep0} in the double-soft limit to an ancillary file attached to this publication, see Appendix~\ref{App:AncillaryFiles}.

\section{Outlook}

The present publication completes the study of one-loop triple-collinear splitting operators and the related splitting functions. The results are sufficient for the construction of an N3LO subtraction scheme. In fact, the only missing input for complete generality is the exact $\epsilon$-dependence of the six-dimensional light-cone-gauge box integral discussed in Section~\ref{Sec:LCGBox}. Should this result be derived in the future, it can be readily substituted in the provided expressions.

For applications at N3LO, there still remains the need to provide the double-soft asymptotics of one-loop matrix-elements squared to sufficient order in the $\epsilon$-expansion. Of course, even with the knowledge of all the limits, the construction of an N3LO subtraction scheme remains a daunting task.

\begin{acknowledgments}
The work of M.C.\ was supported by the Deutsche Forschungsgemeinschaft (DFG) under grant 396021762 - TRR 257: Particle Physics Phenomenology after the Higgs Discovery. The work of S.S.\ was supported by the Polish National Science Centre grant no. 2017/27/B/ST2/02004.
\end{acknowledgments}

\clearpage

\appendix

\section{Definition, representations and expansions of hypergeometric $_2F_1$ functions} \label{App:2F1}

The hypergeometric functions occurring in the expressions of the splitting operators have been chosen to have arguments $0 < x < 1$, which ensures that their defining series converge:
\begin{equation} \label{eq:2F1def}
    \begin{split}
      {}_2F_1(a,b,c,x) &\equiv \sum_{n=0}^\infty \frac{\Gamma(a+n)\Gamma(b+n)\Gamma(c)}{\Gamma(a)\Gamma(b)\Gamma(c+n)} \frac{x^n}{n!} \\[.2cm]
      &= \frac{\Gamma (c)}{\Gamma (b) \Gamma (c-b)}\int_0^1 \dd{t} t^{b-1 }(1-t)^{c-b-1} (1-t x)^{-a} \\[.2cm]
      &= \frac{1}{2\pi i} \int_\mathcal{C} \dd{z} \frac{\Gamma (a+z) \Gamma (b+z) \Gamma (c) \Gamma (-z)}{\Gamma (a) \Gamma (b) \Gamma (c+z)} (-x)^z \\[.2cm]
      &= \frac{1}{2\pi i} \int_\mathcal{C} \dd{z} \frac{\Gamma (a+z) \Gamma (b+z) \Gamma (c) \Gamma (c-a-b-z) \Gamma (-z)}{\Gamma(a) \Gamma (b) \Gamma (c-a) \Gamma (c-b)} (1-x)^z \; .
    \end{split}
\end{equation}
For completeness, we have also recalled the standard ordinary integral representation of the $_2F_1$ function and two of its Mellin-Barnes representations. The integration contours, $\mathcal{C}$, in the latter case should be chosen along the imaginary axis and separate the poles of the $\Gamma$-functions $\Gamma(\dots + z)$ and $\Gamma(\dots - z)$. If this cannot be achieved for the desired value of the dimensional regularisation parameter $\epsilon$, then the correct representation may be obtained with the help of an analytic continuation, for example using the software package \textsc{MB} \cite{Czakon:2005rk}. However, this is not needed for any of the functions listed below. The behaviour of the functions at the endpoint $x = 1$ requires an asymptotic expansion which may be derived with the help of a direct consequence of the integral representation on the second line of Eq.~\eqref{eq:2F1def}:
\begin{equation}
\begin{split}
    _2F_1(a,b,c,x) &= \frac{\Gamma(c-a-b)\Gamma(c)}{\Gamma(c-a)\Gamma(c-b)}\,_2F_1(a,b,a+b-c+1,1-x)\\[.2cm]&\quad+\frac{\Gamma(a+b-c)\Gamma(c)}{\Gamma(a)\Gamma(b)}(1-x)^{c-a-b}\,_2F_1(c-a,c-b,c-a-b+1,1-x) \; .
\end{split}
\end{equation}
The $\epsilon$-expansions of the hypergeometric functions required by the present work may be obtained with the help of the software package \textsc{HypExp} \cite{Maitre:2005uu, Maitre:2007kp, Huber:2005yg}. Below, we reproduce them retaining terms up to transcendental weight six as necessary for applications at N3LO. The results are given in terms of Riemann's $\zeta_n$, classical polylogarithms $\text{Li}_n(x)$ and Nielsen polylogarithms $\text{S}_{n,p}(x)$:
\begin{equation}
    \text{S}_{n,p}(x) \equiv \frac{(-1)^{n+p-1}}{(n-1)!p!} \int_0^1 \dd{t} \frac{\ln^{n-1}(t)\ln^p(1-tx)}{t}\; , \qquad \text{Li}_n(x) = \text{S}_{n-1,1}(x) \; , \qquad \zeta_n = \text{Li}_n(1) \; .
\end{equation}
We also provide expansions at $x = 1$ and, in one case where it is needed, at $x = 0$. The following functions are present in the expressions of master integrals:
\begin{align}
        _2F_1&(-\epsilon,-\epsilon,1-\epsilon,x) = \notag\\[.2cm] &1+\epsilon^2\text{Li}_2(x)+\epsilon^3\Big(-\zeta_3+\text{Li}_3(1-x)+\text{Li}_3(x)-\text{Li}_2(1-x)\ln(1-x)-\frac{1}{2}\ln(x)\ln^2(1-x)\Big)\notag\\[.2cm]&+\epsilon^4\Big(-S_{2,2}(x)+\zeta_4-\text{Li}_4(1-x)+\text{Li}_4(x)-\frac{1}{2}\text{Li}_2(1-x)\ln^2(1-x)+\text{Li}_3(1-x)\ln(1-x)\notag\\[.2cm]&-\frac{1}{6}\ln(x)\ln^3(1-x)\Big)+\epsilon^5\Big(S_{2,3}(x)-S_{3,2}(x)-\zeta_5+\text{Li}_5(1-x)+\text{Li}_5(x)-\frac{1}{6}\text{Li}_2(1-x)\ln^3(1-x)\notag\\[.2cm]&+\frac{1}{2}\text{Li}_3(1-x)\ln^2(1-x)-\text{Li}_4(1-x)\ln(1-x)-\frac{1}{24}\ln(x)\ln^4(1-x)\Big)+\epsilon^6\Big(-S_{2,4}(x)+S_{3,3}(x)\notag\\[.2cm]&-S_{4,2}(x)+\zeta_6-\text{Li}_6(1-x)+\text{Li}_6(x)-\frac{1}{24}\text{Li}_2(1-x)\ln^4(1-x)+\frac{1}{6}\text{Li}_3(1-x)\ln^3(1-x)\notag\\[.2cm]&-\frac{1}{2}\text{Li}_4(1-x)\ln^2(1-x)+\text{Li}_5(1-x)\ln(1-x)-\frac{1}{120}\ln(x)\ln^5(1-x)\Big)+\mathcal{O}\big(\epsilon^7\big) \notag\\[.4cm]
        &= 1+\frac{\epsilon^2}{1-\epsilon}x+\mathcal{O}\big(x^2\big) \notag\\[.4cm]
        &= \Big(\Gamma(1-\epsilon)\Gamma(1+\epsilon)+\mathcal{O}\big(1-x\big)\Big)+(1-x)^{\epsilon}\mathcal{O}\big(1-x\big) \; , \label{eq:2F1case1}
\end{align}
\begin{align}
        _2F_1&(1,1-\epsilon,2-2\epsilon,x) = \notag\\[.2cm]& \frac{1}{x} \Big(-\ln(1-x)+\epsilon\Big(2\text{Li}_2(x)+\frac{1}{2}\ln(1-x)(\ln(1-x)+4)\Big)+\epsilon^2\Big(-2\zeta_3+2\text{Li}_3(1-x)+4\text{Li}_3(x)\notag\\[.2cm]&\quad\;-2\text{Li}_2(1-x)\ln(1-x)-2\text{Li}_2(x)(\ln(1-x)+2)-\frac{1}{6}\ln^3(1-x)-\ln(x)\ln^2(1-x)-\ln^2(1-x)\Big)\notag\\[.2cm]&\quad\;+\epsilon^3\Big(-4S_{2,2}(x)+4\zeta_3+2\zeta_4-4\text{Li}_3(1-x)-8\text{Li}_3(x)-2\text{Li}_4(1-x)+8\text{Li}_4(x)\notag\\[.2cm]&\quad\;+\text{Li}_2(1-x)\ln(1-x)(\ln(1-x)+4)+\text{Li}_2(x)\ln(1-x)(\ln(1-x)+4)-4\text{Li}_3(x)\ln(1-x)\notag\\[.2cm]&\quad\;+2\zeta_3\ln(1-x)+\frac{1}{24}\ln^4(1-x)+\frac{2}{3}\ln(x)\ln^3(1-x)+\frac{1}{3}\ln^3(1-x)+2\ln(x)\ln^2(1-x)\Big)\notag\\[.2cm]&\quad\;+\epsilon^4\Big(8S_{2,2}(x)+4S_{2,3}(x)-8S_{3,2}(x)+4\ln(1-x)S_{2,2}(x)-4\zeta_4-2\zeta_5+4\text{Li}_4(1-x)-16\text{Li}_4(x)\notag\\[.2cm]&\quad\;+2\text{Li}_5(1-x)+16\text{Li}_5(x)-\frac{1}{3}\text{Li}_2(1-x)(\ln(1-x)+6)\ln^2(1-x)\notag\\[.2cm]&\quad\;-\frac{1}{3}\text{Li}_2(x)(\ln(1-x)+6)\ln^2(1-x)+2\text{Li}_3(x)\ln^2(1-x)+8\text{Li}_3(x)\ln(1-x)-8\text{Li}_4(x)\ln(1-x)\notag\\[.2cm]&\quad\;-\zeta_3\ln^2(1-x)-4\zeta_3\ln(1-x)-2\zeta_4\ln(1-x)-\frac{1}{120}\ln^5(1-x)-\frac{1}{4}\ln(x)\ln^4(1-x)\notag\\[.2cm]&\quad\;-\frac{1}{12}\ln^4(1-x)-\frac{4}{3}\ln(x)\ln^3(1-x)\Big)+\epsilon^5\Big(-8S_{2,3}(x)-4S_{2,4}(x)+16S_{3,2}(x)+8S_{3,3}(x)\notag\\[.2cm]&\quad\;-16S_{4,2}(x)-2\ln^2(1-x)S_{2,2}(x)-8\ln(1-x)S_{2,2}(x)-4\ln(1-x)S_{2,3}(x)+8\ln(1-x)S_{3,2}(x)\notag\\[.2cm]&\quad\;+4\zeta_5+2\zeta_6-4\text{Li}_5(1-x)-32\text{Li}_5(x)-2\text{Li}_6(1-x)+32\text{Li}_6(x)\notag\\[.2cm]&\quad\;+\frac{1}{12}\text{Li}_2(1-x)(\ln(1-x)+8)\ln^3(1-x)+\frac{1}{12}\text{Li}_2(x)(\ln(1-x)+8)\ln^3(1-x)\notag\\[.2cm]&\quad\;-\frac{2}{3}\text{Li}_3(x)\ln^3(1-x)-4\text{Li}_3(x)\ln^2(1-x)+4\text{Li}_4(x)\ln^2(1-x)+16\text{Li}_4(x)\ln(1-x)\notag\\[.2cm]&\quad\;-16\text{Li}_5(x)\ln(1-x)+\frac{1}{3}\zeta_3\ln^3(1-x)+2\zeta_3\ln^2(1-x)+\zeta_4\ln^2(1-x)+4\zeta_4\ln(1-x)\notag\\[.2cm]&\quad\;+2\zeta_5\ln(1-x)+\frac{1}{720}\ln^6(1-x)+\frac{1}{15}\ln(x)\ln^5(1-x)+\frac{1}{60}\ln^5(1-x)+\frac{1}{2}\ln(x)\ln^4(1-x)\Big)\notag\\[.2cm]&\quad\;+\mathcal{O}\big(\epsilon^6\big) \Big) \notag\\[.4cm]
        &= \Big(\Big(-\frac{1}{\epsilon}+2\Big)+\mathcal{O}\big(1-x\big)\Big)+(1-x)^{-\epsilon}\Big(\frac{(1-2\epsilon)\Gamma(1-2\epsilon)\Gamma(1+\epsilon)}{\epsilon\Gamma(1-\epsilon)}+\mathcal{O}\big(1-x\big)\Big) \; , \label{eq:2F1case2}
\end{align}
\begin{align}
        _2F_1&(1,1,1-\epsilon,x) = \notag\notag\\[.2cm]&\frac{1}{1-x} \Big( 1-\epsilon\ln(1-x)+\epsilon^2\Big(\text{Li}_2(x)+\frac{1}{2}\ln^2(1-x)\Big)+\epsilon^3\Big(-\zeta_3+\text{Li}_3(1-x)+\text{Li}_3(x)\notag\\[.2cm]&\qquad\quad-\text{Li}_2(1-x)\ln(1-x)-\text{Li}_2(x)\ln(1-x)-\frac{1}{6}\ln^3(1-x)-\frac{1}{2}\ln(x)\ln^2(1-x)\Big)\notag\\[.2cm]&\qquad\quad+\epsilon^4\Big(-S_{2,2}(x)+\zeta_4-\text{Li}_4(1-x)+\text{Li}_4(x)+\frac{1}{2}\text{Li}_2(1-x)\ln^2(1-x)+\frac{1}{2}\text{Li}_2(x)\ln^2(1-x)\notag\\[.2cm]&\qquad\quad-\text{Li}_3(x)\ln(1-x)+\zeta_3\ln(1-x)+\frac{1}{24}\ln^4(1-x)+\frac{1}{3}\ln(x)\ln^3(1-x)\Big)+\epsilon^5\Big(S_{2,3}(x)\notag\\[.2cm]&\qquad\quad-S_{3,2}(x)+\ln(1-x)S_{2,2}(x)-\zeta_5+\text{Li}_5(1-x)+\text{Li}_5(x)-\frac{1}{6}\text{Li}_2(1-x)\ln^3(1-x)\notag\\[.2cm]&\qquad\quad-\frac{1}{6}\text{Li}_2(x)\ln^3(1-x)+\frac{1}{2}\text{Li}_3(x)\ln^2(1-x)-\text{Li}_4(x)\ln(1-x)-\frac{1}{2}\zeta_3\ln^2(1-x)\notag\\[.2cm]&\qquad\quad-\zeta_4\ln(1-x)-\frac{1}{120}\ln^5(1-x)-\frac{1}{8}\ln(x)\ln^4(1-x)\Big)+\epsilon^6\Big(-S_{2,4}(x)+S_{3,3}(x)-S_{4,2}(x)\notag\\[.2cm]&\qquad\quad-\frac{1}{2}\ln^2(1-x)S_{2,2}(x)-\ln(1-x)S_{2,3}(x)+\ln(1-x)S_{3,2}(x)+\zeta_6-\text{Li}_6(1-x)+\text{Li}_6(x)\notag\\[.2cm]&\qquad\quad+\frac{1}{24}\text{Li}_2(1-x)\ln^4(1-x)+\frac{1}{24}\text{Li}_2(x)\ln^4(1-x)-\frac{1}{6}\text{Li}_3(x)\ln^3(1-x)+\frac{1}{2}\text{Li}_4(x)\ln^2(1-x)\notag\\[.2cm]&\qquad\quad-\text{Li}_5(x)\ln(1-x)+\frac{1}{6}\zeta_3\ln^3(1-x)+\frac{1}{2}\zeta_4\ln^2(1-x)+\zeta_5\ln(1-x)+\frac{1}{720}\ln^6(1-x)\notag\\[.2cm]&\qquad\quad+\frac{1}{30}\ln(x)\ln^5(1-x)\Big)+\mathcal{O}\big(\epsilon^7\big) \Big) \notag\\[.4cm]
        &= (1-x)^{-1-\epsilon}\Big(\Gamma(1-\epsilon)\Gamma(1+\epsilon)+\mathcal{O}\big(1-x\big) \Big)+\mathcal{O}\big((1-x)^0\big) \; , \label{eq:2F1case3}
\end{align}
\begin{align}
       _2F_1&(1,1,2+\epsilon,x) = \notag\\[.2cm] &\frac{1}{x} \Big( -\ln(1-x)+\epsilon\Big(-\text{Li}_2(x)-\frac{1}{2}\ln(1-x)(\ln(1-x)+2)\Big)+\epsilon^2\Big(-\zeta_3+\text{Li}_3(1-x)+\text{Li}_3(x)\notag\\[.2cm]&\quad\;-\text{Li}_2(1-x)\ln(1-x)-\text{Li}_2(x)(\ln(1-x)+1)-\frac{1}{6}\ln^3(1-x)-\frac{1}{2}\ln(x)\ln^2(1-x)\notag\\[.2cm]&\quad\;-\frac{1}{2}\ln^2(1-x)\Big)+\epsilon^3\Big(S_{2,2}(x)-\zeta_3-\zeta_4+\text{Li}_3(1-x)+\text{Li}_3(x)+\text{Li}_4(1-x)-\text{Li}_4(x)\notag\\[.2cm]&\quad\;-\frac{1}{2}\text{Li}_2(1-x)\ln(1-x)(\ln(1-x)+2)-\frac{1}{2}\text{Li}_2(x)\ln(1-x)(\ln(1-x)+2)+\text{Li}_3(x)\ln(1-x)\notag\\[.2cm]&\quad\;-\zeta_3\ln(1-x)-\frac{1}{24}\ln^4(1-x)-\frac{1}{3}\ln(x)\ln^3(1-x)-\frac{1}{6}\ln^3(1-x)-\frac{1}{2}\ln(x)\ln^2(1-x)\Big)\notag\\[.2cm]&\quad\;+\epsilon^4\Big(S_{2,2}(x)+S_{2,3}(x)-S_{3,2}(x)+\ln(1-x)S_{2,2}(x)-\zeta_4-\zeta_5+\text{Li}_4(1-x)-\text{Li}_4(x)+\text{Li}_5(1-x)\notag\\[.2cm]&\quad\;+\text{Li}_5(x)-\frac{1}{6}\text{Li}_2(1-x)(\ln(1-x)+3)\ln^2(1-x)-\frac{1}{6}\text{Li}_2(x)(\ln(1-x)+3)\ln^2(1-x)\notag\\[.2cm]&\quad\;+\frac{1}{2}\text{Li}_3(x)\ln^2(1-x)+\text{Li}_3(x)\ln(1-x)-\text{Li}_4(x)\ln(1-x)-\frac{1}{2}\zeta_3\ln^2(1-x)-\zeta_3\ln(1-x)\notag\\[.2cm]&\quad\;-\zeta_4\ln(1-x)-\frac{1}{120}\ln^5(1-x)-\frac{1}{8}\ln(x)\ln^4(1-x)-\frac{1}{24}\ln^4(1-x)-\frac{1}{3}\ln(x)\ln^3(1-x)\Big)\notag\\[.2cm]&\quad\;+\epsilon^5\Big(S_{2,3}(x)+S_{2,4}(x)-S_{3,2}(x)-S_{3,3}(x)+S_{4,2}(x)+\frac{1}{2}\ln^2(1-x)S_{2,2}(x)+\ln(1-x)S_{2,2}(x)\notag\\[.2cm]&\quad\;+\ln(1-x)S_{2,3}(x)-\ln(1-x)S_{3,2}(x)-\zeta_5-\zeta_6+\text{Li}_5(1-x)+\text{Li}_5(x)+\text{Li}_6(1-x)-\text{Li}_6(x)\notag\\[.2cm]&\quad\;-\frac{1}{24}\text{Li}_2(1-x)(\ln(1-x)+4)\ln^3(1-x)-\frac{1}{24}\text{Li}_2(x)(\ln(1-x)+4)\ln^3(1-x)\notag\\[.2cm]&\quad\;+\frac{1}{6}\text{Li}_3(x)\ln^3(1-x)+\frac{1}{2}\text{Li}_3(x)\ln^2(1-x)-\frac{1}{2}\text{Li}_4(x)\ln^2(1-x)-\text{Li}_4(x)\ln(1-x)\notag\\[.2cm]&\quad\;+\text{Li}_5(x)\ln(1-x)-\frac{1}{6}\zeta_3\ln^3(1-x)-\frac{1}{2}\zeta_3\ln^2(1-x)-\frac{1}{2}\zeta_4\ln^2(1-x)-\zeta_4\ln(1-x)\notag\\[.2cm]&\quad\;-\zeta_5\ln(1-x)-\frac{1}{720}\ln^6(1-x)-\frac{1}{30}\ln(x)\ln^5(1-x)-\frac{1}{120}\ln^5(1-x)-\frac{1}{8}\ln(x)\ln^4(1-x)\Big)\notag\\[.2cm]&\quad\;+\mathcal{O}\big(\epsilon^6\big) \Big) \notag\\[.4cm]
       &= \Big(\Big(\frac{1}{\epsilon}+1\Big)+\mathcal{O}\big(1-x\big)\Big)+(1-x)^{\epsilon}\Big(-\frac{(1+\epsilon)\Gamma(1-\epsilon)\Gamma(1+\epsilon)}{\epsilon}+\mathcal{O}\big(1-x\big)\Big) \; . \label{eq:2F1case4}
\end{align}
The following functions are either special cases or occur in the asymptotic behaviour of the Appell $F_1$ function discussed in Appendix~\ref{App:F1}:
\begin{align}
        _2F_1&(1,1-\epsilon,1-2\epsilon,x) = \notag\\[.2cm]& \frac{1}{1-x} \Big(1-\epsilon\ln(1-x)+\epsilon^2\Big(2\text{Li}_2(x)+\frac{1}{2}\ln^2(1-x)\Big)+\epsilon^3\Big(-2\zeta_3+2\text{Li}_3(1-x)+4\text{Li}_3(x)\notag\\[.2cm]&\qquad\quad-2\text{Li}_2(1-x)\ln(1-x)-2\text{Li}_2(x)\ln(1-x)-\frac{1}{6}\ln^3(1-x)-\ln(x)\ln^2(1-x)\Big)\notag\\[.2cm]&\qquad\quad+\epsilon^4\Big(-4S_{2,2}(x)+2\zeta_4-2\text{Li}_4(1-x)+8\text{Li}_4(x)+\text{Li}_2(1-x)\ln^2(1-x)+\text{Li}_2(x)\ln^2(1-x)\notag\\[.2cm]&\qquad\quad-4\text{Li}_3(x)\ln(1-x)+2\zeta_3\ln(1-x)+\frac{1}{24}\ln^4(1-x)+\frac{2}{3}\ln(x)\ln^3(1-x)\Big)+\epsilon^5\Big(4S_{2,3}(x)\notag\\[.2cm]&\qquad\quad-8S_{3,2}(x)+4\ln(1-x)S_{2,2}(x)-2\zeta_5+2\text{Li}_5(1-x)+16\text{Li}_5(x)\notag\\[.2cm]&\qquad\quad-\frac{1}{3}\text{Li}_2(1-x)\ln^3(1-x)-\frac{1}{3}\text{Li}_2(x)\ln^3(1-x)+2\text{Li}_3(x)\ln^2(1-x)-8\text{Li}_4(x)\ln(1-x)\notag\\[.2cm]&\qquad\quad-\zeta_3\ln^2(1-x)-2\zeta_4\ln(1-x)-\frac{1}{120}\ln^5(1-x)-\frac{1}{4}\ln(x)\ln^4(1-x)\Big)+\epsilon^6\Big(-4S_{2,4}(x)\notag\\[.2cm]&\qquad\quad+8S_{3,3}(x)-16S_{4,2}(x)-2\ln^2(1-x)S_{2,2}(x)-4\ln(1-x)S_{2,3}(x)+8\ln(1-x)S_{3,2}(x)\notag\\[.2cm]&\qquad\quad+2\zeta_6-2\text{Li}_6(1-x)+32\text{Li}_6(x)+\frac{1}{12}\text{Li}_ 2(1-x)\ln^4(1-x)+\frac{1}{12}\text{Li}_2(x)\ln^4(1-x)\notag\\[.2cm]&\qquad\quad-\frac{2}{3}\text{Li}_3(x)\ln^3(1-x)+4\text{Li}_4(x)\ln^2(1-x)-16\text{Li}_5(x)\ln(1-x)+\frac{1}{3}\zeta_3\ln^3(1-x)\notag\\[.2cm]&\qquad\quad+\zeta_4\ln^2(1-x)+2\zeta_5\ln(1-x)+\frac{1}{720}\ln^6(1-x)+\frac{1}{15}\ln(x)\ln^5(1-x)\Big)+\mathcal{O}\big(\epsilon^7\big) \Big) \notag\\[.4cm] &= (1-x)^{-1-\epsilon}\Big(\frac{\Gamma(1-2\epsilon)\Gamma(1+\epsilon)}{\Gamma(1-\epsilon)}+\mathcal{O}\big(1-x\big)\Big)+\mathcal{O}\big((1-x)^0\big) \; , \label{eq:F1case1}
\end{align}
\begin{align}
        _2F_1&(1,-\epsilon,1-2\epsilon,x) = \notag\\[.2cm]& 1+\epsilon\ln(1-x)+\epsilon^2\Big(-2\text{Li}_2(x)-\frac{1}{2}\ln^2(1-x)\Big)+\epsilon^3\Big(2\zeta_3-2\text{Li}_3(1-x)-4\text{Li}_3(x)\notag\\[.2cm]&+2\text{Li}_2(1-x)\ln(1-x)+2\text{Li}_2(x)\ln(1-x)+\frac{1}{6}\ln^3(1-x)+\ln(x)\ln^2(1-x)\Big)+\epsilon^4\Big(4S_{2,2}(x)\notag\\[.2cm]&-2\zeta_4+2\text{Li}_4(1-x)-8\text{Li}_4(x)-\text{Li}_2(1-x)\ln^2(1-x)-\text{Li}_2(x)\ln^2(1-x)+4\text{Li}_3(x)\ln(1-x)\notag\\[.2cm]&-2\zeta_3\ln(1-x)-\frac{1}{24}\ln^4(1-x)-\frac{2}{3}\ln(x)\ln^3(1-x)\Big)+\epsilon^5\Big(-4S_{2,3}(x)+8S_{3,2}(x)\notag\\[.2cm]&-4\ln(1-x)S_{2,2}(x)+2\zeta_5-2\text{Li}_5(1-x)-16\text{Li}_5(x)+\frac{1}{3}\text{Li}_2(1-x)\ln^3(1-x)+\frac{1}{3}\text{Li}_2(x)\ln^3(1-x)\notag\\[.2cm]&-2\text{Li}_3(x)\ln^2(1-x)+8\text{Li}_4(x)\ln(1-x)+\zeta_3\ln^2(1-x)+2\zeta_4\ln(1-x)+\frac{1}{120}\ln^5(1-x)\notag\\[.2cm]&+\frac{1}{4}\ln(x)\ln^4(1-x)\Big)+\epsilon^6\Big(4S_{2,4}(x)-8S_{3,3}(x)+16S_{4,2}(x)+2\ln^2(1-x)S_{2,2}(x)\notag\\[.2cm]&+4\ln(1-x)S_{2,3}(x)-8\ln(1-x)S_{3,2}(x)-2\zeta_6+2\text{Li}_6(1-x)-32\text{Li}_6(x)-\frac{1}{12}\text{Li}_2(1-x)\ln^4(1-x)\notag\\[.2cm]&-\frac{1}{12}\text{Li}_2(x)\ln^4(1-x)+\frac{2}{3}\text{Li}_3(x)\ln^3(1-x)-4\text{Li}_4(x)\ln^2(1-x)+16\text{Li}_5(x)\ln(1-x)\notag\\[.2cm]&-\frac{1}{3}\zeta_3\ln^3(1-x)-\zeta_4\ln^2(1-x)-2\zeta_5\ln(1-x)-\frac{1}{720}\ln^6(1-x)-\frac{1}{15}\ln(x)\ln^5(1-x)\Big)+\mathcal{O}\big(\epsilon^7\big) \notag\\[.4cm] &= \Big(2+\mathcal{O}\big(1-x\big)\Big)+(1-x)^{-\epsilon}\Big(-\frac{\Gamma(1-2\epsilon)\Gamma(1+\epsilon)}{\Gamma(1-\epsilon)}+\mathcal{O}\big(1-x\big)\Big) \; , \label{eq:F1case2}
\end{align}
\begin{align}
        _2F_1&(-\epsilon,1+\epsilon,1-\epsilon,x) = \notag\\[.2cm]&
        1+\epsilon\ln(1-x)+\epsilon^2\Big(-2\text{Li}_2(x)-\frac{1}{2}\ln^2(1-x)\Big)+\epsilon^3\Big(-2\zeta_3+2\text{Li}_3(1-x)-2\text{Li}_3(x)\notag\\[.2cm]&-2\text{Li}_2(1-x)\ln(1-x)+\frac{1}{6}\ln^3(1-x)-\ln(x)\ln^2(1-x)\Big)+\epsilon^4\Big(-2S_{2,2}(x)-2\zeta_4+2\text{Li}_4(1-x)\notag\\[.2cm]&-2\text{Li}_4(x)+\text{Li}_2(1-x)\ln^2(1-x)-2\text{Li}_3(1-x)\ln(1-x)-\frac{1}{24}\ln^4(1-x)+\frac{1}{3}\ln(x)\ln^3(1-x)\Big)\notag\\[.2cm]&+\epsilon^5\Big(-2S_{2,3}(x)-2S_{3,2}(x)-2\zeta_5+2\text{Li}_5(1-x)-2\text{Li}_5(x)-\frac{1}{3}\text{Li}_2(1-x)\ln^3(1-x)\notag\\[.2cm]&+\text{Li}_3(1-x)\ln^2(1-x)-2\text{Li}_4(1-x)\ln(1-x)+\frac{1}{120}\ln^5(1-x)-\frac{1}{12}\ln(x)\ln^4(1-x)\Big)\notag\\[.2cm]&+\epsilon^6\Big(-2S_{2,4}(x)-2S_{3,3}(x)-2S_{4,2}(x)-2\zeta_6+2\text{Li}_6(1-x)-2\text{Li}_6(x)+\frac{1}{12}\text{Li}_2(1-x)\ln^4(1-x)\notag\\[.2cm]&-\frac{1}{3}\text{Li}_3(1-x)\ln^3(1-x)+\text{Li}_4(1-x)\ln^2(1-x)-2\text{Li}_5(1-x)\ln(1-x)-\frac{1}{720}\ln^6(1-x)\notag\\[.2cm]&+\frac{1}{60}\ln(x)\ln^5(1-x)\Big)+\mathcal{O}\big(\epsilon^7\big) \notag\\[.4cm]
        &= \Big(\frac{2\Gamma(1-\epsilon)^2}{\Gamma(1-2\epsilon)}+\mathcal{O}\big(1-x\big)\Big)+(1-x)^{-\epsilon}\Big(-1+\mathcal{O}\big(1-x\big)\Big) \; . \label{eq:F1case3}
\end{align}

\section{Definition, representations and expansions of the Appell $F_1$ function} \label{App:F1}

The Appell $F_1$ function is defined as follows:
\begin{equation} \label{eq:F1def}
    \begin{split}
      F_1&(a,b_1,b_2,c,x,y) \equiv \sum_{m=0}^\infty \sum_{n=0}^\infty \frac{\Gamma(a+m+n)\Gamma(b_1+m)\Gamma(b_2+n)\Gamma(c)}{\Gamma(a)\Gamma(b_1)\Gamma(b_2)\Gamma(c+m+n)} \frac{x^m}{m!} \frac{y^n}{n!}\\[.4cm]
      &= \frac{\Gamma (c)}{\Gamma (a) \Gamma (c-a)}\int_0^1 \dd{t} t^{a-1}(1-t)^{c-a-1} (1-t x)^{-b_1} (1-t y)^{-b_2} \\[.4cm]
      &= \frac{1}{(2\pi i)^2} \iint_\mathcal{C} \dd{z_1} \dd{z_2} \frac{\Gamma(a+z_1+z_2)\Gamma(b_1+z_1)\Gamma(b_2+z_2)\Gamma(c)\Gamma(-z_1)\Gamma(-z_2)}{\Gamma(a)\Gamma(b_1)\Gamma(b_2)\Gamma(c+z_1+z_2)}(-x)^{z_1}(-y)^{z_2} \\[.4cm]
      &= \frac{1}{(2\pi i)^2} \iint_\mathcal{C} \dd{z_1} \dd{z_2} \frac{\Gamma(a+z_1+z_2)\Gamma(b_1+z_1)\Gamma(b_2+z_2)\Gamma(c-a-b_1-z_1)\Gamma(c)}{\Gamma(a)\Gamma(b_1)\Gamma(b_2)\Gamma(c-a)\Gamma(c-b_1+z_2)} \\[.2cm]
      &\qquad\qquad\qquad\qquad \times \Gamma(-z_1)\Gamma(-z_2)(1-x)^{z_1}(-y)^{z_2} \\[.4cm]
      &= \frac{1}{(2\pi i)^2} \iint_\mathcal{C} \dd{z_1} \dd{z_2} \frac{\Gamma(a+z_1+z_2)\Gamma(b_1+z_1)\Gamma(b_2+z_2)\Gamma(c-a-b_1-b_2-z_1-z_2)\Gamma(c)}{\Gamma(a)\Gamma(b_1)\Gamma(b_2)\Gamma(c-a)\Gamma(c-b_1-b_2)} \\[.2cm]
      &\qquad\qquad\qquad\qquad \times \Gamma(-z_1)\Gamma(-z_2)(1-x)^{z_1}(1-y)^{z_2} \; .
    \end{split}
\end{equation}
As in the case of the hypergeometric function Appendix~\ref{App:2F1}, the integration contours of the Mellin-Barnes representations starting with the third equality, $\mathcal{C}$, should be chosen along the imaginary axis and separate the poles of the $\Gamma$-functions $\Gamma(\dots + z)$ and $\Gamma(\dots - z)$. It turns out that only one Appell function is actually needed in the present case. We choose it to have arguments $0 < x \leq y < 1$ ensuring the convergence of the series representation Eq.~\eqref{eq:F1def}:
\begin{equation} \label{eq:F1exp}
    \begin{split}
        F_1(1&,-\epsilon,1,1-2\epsilon,1-x,1-y) = \\[.2cm] &\frac{1}{y} \Big( 1+\epsilon(\ln(x)-2\ln(y))+\frac{1}{2}\epsilon^2\Big(4\text{Li}_2\Big(\frac{x}{y}\Big)-4\text{Li}_2(y)-4\ln(x)\ln(y)-4\ln(y)\ln(y-x)\\[.2cm]&\quad+4\ln(x)\ln(y-x)-\ln^2(x)+6\ln^2(y)-4\ln(1-y)\ln(y)\Big)+\frac{1}{6}\epsilon^3\Big(-12\zeta_3-12\text{Li}_3\Big(1-\frac{1-y}{1-x}\Big)\\[.2cm]&\quad-24\text{Li}_3\Big(1-\frac{x}{y}\Big)+12\text{Li}_3\Big(1-\frac{x(1-y)}{(1-x)y}\Big)-12\text{Li}_3\Big(\frac{1-y}{1-x}\Big)+12\text{Li}_3\Big(\frac{x(1-y)}{(1-x)y}\Big)\\[.2cm]&\quad-12\ln(x)\text{Li}_2\Big(\frac{1-y}{1-x}\Big)-12\ln(x)\text{Li}_2\Big(\frac{x}{y}\Big)-12\text{Li}_2(y)\ln(x)-12\text{Li}_3(x)+12\text{Li}_2(x)\ln(x)\\[.2cm]&\quad+24\text{Li}_3(1-y)+24\text{Li}_3(y)+24\zeta_2\ln(x)-12\ln^2(1-x)\ln(y)+6\ln^2(x)\ln(y)\\[.2cm]&\quad-6\ln^2(x)\ln(y-x)-12\ln(1-x)\ln^2(y)+6\ln^2(y)\ln(y-x)\\[.2cm]&\quad+12\ln(1-x)\ln(x)\ln(y)+12\ln(1-x)\ln(1-y)\ln(y)-12\ln(x)\ln(1-y)\ln(y)\\[.2cm]&\quad+12\ln(1-x)\ln(y)\ln(y-x)-12\ln(1-y)\ln(y)\ln(y-x)+\ln^3(x)-12\zeta_2\ln(y)-8\ln^3(y)\\[.2cm]&\quad+18\ln(1-y)\ln^2(y)\Big)+\mathcal{O}\big(\epsilon^4\big) \Big) \; .
    \end{split}
\end{equation}
The above result has been obtained with the software package \textsc{PolyLogTools} \cite{Duhr:2019tlz} by direct integration of the integral representation after remapping the integration region from $[0,1]$ to $[0,\infty)$. The series expansion has been truncated after enough terms to match the requirements of N3LO applications as discussed in the main part of the text. Care has been taken, so that the arguments of all functions belong to the unit interval. The following special cases are also required with deeper $\epsilon$-expansions:
\begin{equation}
    \begin{aligned}
        \text{F}_1(1,-\epsilon,1,1-2\epsilon,1-x,1-x) &= {}_2F_1(1,1-\epsilon,1-2\epsilon,1-x) \; , \\[.4cm]
        \text{F}_1(1,-\epsilon,1,1-2\epsilon,1-x,0) &= {}_2F_1(1,-\epsilon,1-2\epsilon,1-x) \; .
    \end{aligned}
\end{equation}
The corresponding $\epsilon$-expansions of the hypergeometric functions are given in Eqs.~\eqref{eq:F1case1} and \eqref{eq:F1case2}, while the above expressions are a direct consequence of the integral representations in Eqs.~\eqref{eq:F1def} and \eqref{eq:2F1def}. Finally, the following asymptotics of the Appell $F_1$ function are needed as well:
\begin{equation} \label{eq:F1asy}
\begin{split}
    \text{F}_1(1,-\epsilon,1,1-2\epsilon,1-x,1-y) \; &\sim \; 2\,_2F_1(1,1,1-\epsilon,1-y)-\frac{x^{-\epsilon}\Gamma(1-2\epsilon)\Gamma(1+\epsilon)}{y\Gamma(1-\epsilon)} \quad \big( x \to 0 \big) \; , \\[.4cm]
    \text{F}_1(1,-\epsilon,1,1-2\epsilon,1-x,1-y) \; &\sim \; \frac{y^{-1-\epsilon}\Gamma(1-2\epsilon)\Gamma(1+\epsilon)}{\Gamma(1-\epsilon)} \,_2F_1\Big(-\epsilon,1+\epsilon,1-\epsilon,1-\frac{x}{y}\Big) \quad \big( y \to 0 \big) \; .
\end{split}
\end{equation}
These results can be obtained by an asymptotic expansion of the Mellin-Barnes representations Eq.~\eqref{eq:F1def}, while the two hypergeometric functions are given in Eqs.~\eqref{eq:2F1case3} and \eqref{eq:F1case3}.

\section{Description of the results provided in ancillary files} \label{App:AncillaryFiles}

The following files are provided:
\begin{itemize}
\item \texttt{integrals.m}, \texttt{expandedIntegrals.m} - substitution lists for the exact (\texttt{integrals}) and expanded up to $\order{\epsilon^0}$ (\texttt{expandedIntegrals}) master integrals, Eqs.~\eqref{eq:Bubble1}, \eqref{eq:Bubble2}, \eqref{eq:OffShellBox}, \eqref{eq:LCGTriangleBubble}, \eqref{eq:LCGOffShellTriangle1}, \eqref{eq:LCGOffShellTriangle2} and \eqref{eq:LCGBoxTriangle}, divided by $r_\Gamma$, Eq.~\eqref{eq:rGamma}, including the results with required permutations of the external momenta; the notation for the integrals in the substitutions matches the occurrences in the files containing the splitting operators and splitting functions listed below; the result for the light-cone-gauge six-dimensional box integral of Section~\ref{Sec:LCGBox} is not provided here but rather substituted as \texttt{LCGBox[pi,pj,pk,d+2]}, where \texttt{pi,pj,pk} correspond to a permutation of $p_1, p_2$ and $p_3$ for which the result is provided in Eq.~\eqref{eq:I11111ep0} and in the file \texttt{LCGBox.m};
\item \texttt{AppellF1.m} - substitution giving the $\epsilon$-expansion of the Appell function Eq.~\eqref{eq:F1exp};
\item \texttt{LCGBox.m}, \texttt{LCGBoxDoubleSoft.m} - right-hand side of Eq.~\eqref{eq:I11111ep0} named \texttt{LCGBoxEp0}, and its double-soft limit at $\order{\epsilon^0}$ (\texttt{LCGBoxDoubleSoftEp0}) and $\order{\epsilon}$ (\texttt{LCGBoxDoubleSoftEp1});
\item \texttt{Pggg.m}, \texttt{Pgqqbar.m}, \texttt{Pqgg.m}, \texttt{Pqqqbar.m}, \texttt{Pqqpqpbar.m} - exact results at tree and one-loop level, the latter expressed through the master integrals, for the splitting (\texttt{P0Pol} and \texttt{P1Pol}) and averaged splitting (\texttt{P0Avg} and \texttt{P1Avg}) functions, Eq.~\eqref{eq:SplittingFunctions}, for the processes $g \to ggg$, $g \to gq\bar{q}$, $q \to qgg$, $q \to qq\bar{q}$ and $q \to qq'\bar{q}'$ respectively; the tree-level results, \texttt{P0Pol} and \texttt{P0Avg}, additionally contain the singularities, proportional to the tag \texttt{Ioperator}, of the one-loop splitting and averaged splitting functions obtained using Eq.~\eqref{eq:Ioperator} that has been partially expanded in $\epsilon$ as described in the text after the equation; the one-loop level results, \texttt{P1Pol} and \texttt{P1Avg}, defined in Eq.~\eqref{eq:SplittingFunctions1l}, are given without the hermitian conjugate;
\item \texttt{P1gggExp.m}, \texttt{P1gqqbarExp.m}, \texttt{P1qggExp.m}, \texttt{P1qqqbarExp.m}, \texttt{P1qqpqpbarExp.m} - one-loop splitting (\texttt{P1PolExp}) and averaged splitting (\texttt{P1AvgExp}) functions from the files listed in the previous item, expanded up to $\order{\epsilon^0}$ after substitution of the master integrals from \texttt{expandedIntegrals}, which implies that they are divided by $r_\Gamma$ as well; 
\item \texttt{ggg0l.m}, \texttt{gqqbar0l.m}, \texttt{qgg0l.m}, \texttt{qqqbar0l.m}, \texttt{qqpqpbar0l.m}, \texttt{ggg1l.m}, \texttt{gqqbar1l.m}, \texttt{qgg1l.m}, \texttt{qqqbar1l.m}, \texttt{qqpqpbar1l.m} - exact splitting operators at tree (\texttt{*0l.m}) and one-loop (\texttt{*1l.m}) level, the latter expressed through the master integrals; each operator has been multiplied with $1/2\,s_{123}$ to make it dimensionless and easier to relate to the respective splitting function; the operators are named correspondingly to the file name (without the extension \texttt{.m} of course); the tree-level results contain the singularities, tagged with the factor \texttt{Ioperator}, of the one-loop splitting operators obtained using Eq.~\eqref{eq:Ioperator} that has partially been expanded in $\epsilon$ as described in the text after the equation;
\item \texttt{gqqbar0l4D.m}, \texttt{qgg0l4D.m}, \texttt{qqqbar0l4D.m}, \texttt{qqpqpbar0l4D.m}, \texttt{gqqbar1l4D.m}, \texttt{qgg1l4D.m}, \\ \texttt{qqqbar1l4D.m}, \texttt{qqpqpbar1l4D.m} - four-dimensional projections, obtained using Eq.~\eqref{eq:4Dprojection}, of the results listed in the previous item at tree and one-loop level; the projections of the pure-gluon splitting operators are not provided, since they are lengthier than the original expressions and lack usefulness;
\item \texttt{README} - notation used in the files.
\end{itemize}
Although the splitting operators and splitting functions are not renormalised, subtracting from them the singularities given in Eq.~\eqref{eq:Ioperator} as provided in the above files, gives the expressions needed to correctly subtract the triple-collinear limit of one-loop amplitudes according to Eq.~\eqref{eq:TripleCollinearSubtraction}, in the `t Hooft-Veltman scheme (four-dimensional external gluon polarisation vectors, but $d$-dimensional internal virtual gluon fields). These expressions thus match the default conventions of the majority of software one-loop amplitude providers, e.g.\ \textsc{NJet} \cite{Badger:2012pg}.

\newpage

\bibliographystyle{JHEP}

\providecommand{\href}[2]{#2}\begingroup\raggedright\endgroup

\end{document}